\documentclass[fleqn,usenatbib]{mnras}

\usepackage{float}
\usepackage{amssymb, amsmath, epsfig}
\usepackage{graphics,times,array,ctable,amsmath,amssymb,natbib}
\usepackage{threeparttable}
\newcommand{\HI}{H{\sc \ i}}

\newcommand{\HeII}{He{\sc \, ii}}
\newcommand{\HeIII}{He{\sc \, iii}}

\def\apj{ApJ}
\def\apjl{ApJL}

\def\aj{AJ}

\def\mnras{MNRAS}

\def\aap{{A\&A}}

\title[\HeII\ reionization in Simulations]{Inhomogeneous \HeII\ reionization in Hydrodynamic Simulations}
\author[Upton Sanderbeck and Bird]{Phoebe Upton Sanderbeck,$^{1}$\thanks{Email: phoebeu@ucr.edu}
Simeon Bird$^{1}$
\\
$^1$Department of Physics \& Astronomy, University of California, Riverside}

\begin{document}
\label{firstpage}
\pagerange{\pageref{firstpage}--\pageref{lastpage}}
\maketitle

\begin{abstract}
The reionization of the second electron of helium shapes the physical state of intergalactic gas at redshifts between $2 \lesssim z \lesssim5$. Because performing full in situ radiative transfer in hydrodynamic simulations is computationally expensive for large volumes, the physics of \HeII\ reionization is often approximated by a uniform UV background model that does not capture the spatial inhomogeneity of reionization. We have devised a model that implements the effects of \HeII\ reionization using semi-analytic calculations of the thermal state of intergalactic gas-- a way to bypass a full radiative transfer simulation while still realizing the physics of \HeII\ reionization that affects observables such as the Lyman $\alpha$ forest. Here we present a publicly-available code that flexibly models inhomogeneous \HeII\ reionization in simulations at a negligible computational cost. Because many of the parameters of \HeII\ reionization are uncertain, our model is customizable from a set of free parameters. We show results from this code in {\sc MP-Gadget}, where this model is implemented. We demonstrate the resulting temperature evolution and temperature-density relation of intergalactic gas-- consistent with recent measurements and previous radiative transfer simulations. We show that the impact of \HeII\ reionization gives rise to subtle signatures in the one-dimensional statistics of the Lyman $\alpha$ forest at the level of several percent, in agreement with previous findings. The flexible nature of these simulations is ideal for studies of \HeII\ reionization and future observations of the \HeII\ Lyman $\alpha$ forest.

\end{abstract}


\section{Introduction}
\label{sec:intro}
The reionization of the second electron of helium is believed to be the last major heating event in of the bulk of the Universe's baryonic matter, the intergalactic medium (IGM). While the reionization of hydrogen and the first electron of helium was likely seeded by the first galaxies and stars, helium was likely not fully ionized until photons of a higher energy regime ($E_{\gamma}\geq4$ Rydberg) were produced in bulk \cite[e.g.][]{miraldaescude93,madau94,miraldaescude00,schaye00,mcquinn09,compostella13,compostella14,laplante17,laplante18}. 

This reionization of the second electron of helium, \HeII\ reionization, was likely driven by quasars. Firstly, the opacity in the \HeII\ Lyman $\alpha$ forest suggests the existence of $\sim10$ comoving Mpc fluctuations that are most plausibly sourced by AGN. Secondly, the observed luminosity function and spectral energy distribution of quasars produces enough 4 Ryd photons to completely reionize \HeII\ by $z\sim3$ \cite[e.g.][]{madau99,miraldaescude00,wyithe03,furlanetto08,haardt12,khaire17,puchwein19,kulkarni19} and Gunn-Peterson troughs in the \HeII\ Lyman-$\alpha$ forest indicate that \HeII\ reionization ended around this time \citep{miraldaescude93,reimers97,hogan97,heap00,zheng04,mcquinnGP, shull10,worseck11, worseck19}. While sources other than quasars emit 4 Rydberg photons (e.g. hot virialized gas, hot galactic and circumgalactic gas, and X-ray binaries), it is unlikely that any of these sources would produce a significant fraction of the extragalactic \HeII-ionizing radiation \citep{uptonsanderbeck18}.

\HeII\ reionization shapes the thermal state of the IGM at $2 \lesssim z \lesssim 5$, photoheating the gas by up to tens of thousands of degrees Kelvin \cite[e.g.][]{miraldaescude94,schaye00,tittley07,mcquinn09,compostella13,compostella14,puchwein15,uptonsanderbeck16}. Because \HeII\ reionization is a patchy process, with different regions becoming reionized at different times, it induces spatial fluctuations in temperature at correlated scales of ten of comoving Mpc, surviving for approximately a Hubble time \citep{miraldaescude94,gnedin97,hui03}. A secondary effect of the thermal impact of reionization, pressure smoothing, alters the spatial distribution of intergalactic gas as well as the subsequent thermal evolution. Both the thermal state of the gas and resulting pressure smoothing affect Lyman $\alpha$ forest absorption-- the most widely used observable of the IGM at $2 \lesssim z \lesssim 5$.

Furthermore, the thermal impact of \HeII\ reionization has a dramatic effect on the temperature-density relation of intergalactic gas. Following the end of photoheating from reionization, the temperature-density relation of the low-density IGM begins settles into an approximate single power-law relation \citep{gnedin97,mcquinn16}. Not only does \HeII\ reionization modulate the slope of the power law, but it also introduces a significant amount of scatter as a consequence of the patchiness \HeII\ reionization. Accurately characterizing the temperature-density relation of the IGM by implementing \HeII\ reionization into Lyman $\alpha$ forest simulations may reduce biases from cosmological parameters extracted from the Lyman $\alpha$ forest.

Many have forecasted the impact of temperature fluctuations on several IGM statistics. \citet{lai06} calculated the effect of temperature fluctuations from a patchy and extended reionization on the Lyman $\alpha$ forest one-dimensional flux power spectrum. They found that models with order unity temperature fluctuations at scales of tens of comoving Mpc only had a few-percent level effect on the flux power spectrum for small scales. Though \citet{mcquinn11} reaffirmed the small effect of \HeII\ reionization on such one-dimensional Lyman $\alpha$ forest statistics, they show that the three dimensional flux power spectrum could show order unity deviation from a model without temperature fluctuations. However, \citet{laplante17} found that inhomogeneities from \HeII\ reionization can induce several percent level effects in the one dimensional flux power spectrum; and in \citet{onorbe19}, the relic thermal fluctuations from hydrogen reionization are found to increase power by up to fifty percent at large scales ($k\lesssim 0.1$ (km/s)$^{-1}$).

Additionally, \citet{meiksin00}, \citet{theuns00}, and \citet{zaldarriaga02} argue that wavelet filters used on Lyman $\alpha$ forest spectra can effectively detect such spatial temperature fluctuations. Two- and three- point statistics from many Lyman $\alpha$ forest sightlines may also be sensitive to \HeII\ reionization temperature fluctuations \citep{white10}, with a three-point function also being able to detect scatter in the temperature-density relation \citep{fang04}.

Generally, \HeII\ reionization in simulations without in situ radiative transfer is often treated using a spatially-uniform ionizing background model that can recreate the mean thermal history at the mean cosmic density owing to \HeII\ reionization. Some of the most widely used ionizing background models in such simulations include \citet{haardt12}, \citet{puchwein15}, \citet{puchwein19}, and \citet{fg19}. The approach of using a uniform ultraviolet background model to capture the effects of reionization neglects the patchiness and resulting fluctuations in temperature, ionization state, and pressure smoothing. This also leads to an artificially narrow temperature-density relation that does not capture the scatter that should result from the inhomogeneity of \HeII\ reionization.

In this paper, we present a method of implementing the thermal effects of \HeII\ reionization into hydrodynamic simulations without the need for an in situ radiative transfer treatment. While radiative transfer simulations can be an order of magnitude more computationally expensive than non-radiative simulations, our model incurs a negligible computational cost. We alter the thermal state of the gas to mimic the radiative transfer, modeling the impact of \HeII\ reionization with a small set of parameters. This code is publicly available as a feature implemented within {\sc MP-Gadget}\footnote{\url{https://github.com/MP-Gadget/MP-Gadget}}.

Our method captures the patchiness of \HeII\ reionization-- an important feature for cosmological simulations and for investigating large scales where it is measurable. The morphology of \HeII\ reionization in our simulations approximates that found in the radiative transfer \HeII\ reionization simulations of \citet{mcquinn09} where large \HeIII\ bubbles form rapidly-- emerging faster than they grow-- and overlap until they fill the entire volume.

 In Section~\ref{sec:model}, we describe the physics of \HeII\ reionization and the free parameters of our model. Section~\ref{sec:results} describes the details and results of our simulations. We conclude in Section~\ref{sec:conclusions}.
 
\section{\HeII\ reionization model}

\label{sec:model}

\subsection{Thermal physics model}
\label{sec:thermal}
The reionization of \HeII\ has a distinct morphology from hydrogen reionization, as it is accomplished by a different class of sources and photon energies. The higher energy photons produced by quasars have long mean free paths that do not collectively form ionization fronts \citep{mcquinn09}. Approximately, \HeII\ -ionizing photons can be categorized into two regimes -- short mean free path photons that nearly instantaneously ionize the quasar's surrounding medium and drive the production and growth of \HeIII\ bubbles, and long mean free path photons that free stream through the Universe and uniformly heat the gas through their rare photoionizations. This multi-zone heating is found in detailed simulations of \HeII\ reionization \citep{mcquinn09} and serves as the basis of analytic models of the thermal history \citep{uptonsanderbeck16}.
 
The determination of whether any given photon will contribute to the formation and growth of a \HeIII\ bubble, or free stream and ultimately uniformly heat the IGM can be roughly distinguished by considering the photon's mean free path. The mean free path can be estimated by the following,

 \begin{equation}
\lambda_{\rm MFP, HeII}\approx5\,\bar{x}^{-1}_{\rm HeII }\left(\frac{E_{\gamma}}{100\rm eV}\right)^3\left(\frac{1+z}{4}\right)^{-2} {\rm cMpc},
\label{eqn:mfp}
\end{equation}
where $E_\gamma$ is the energy of the photon. If the mean free path of the photon is greater than the typical size of a \HeIII\ bubble, then this photon will contribute to a more uniform background of these higher energy photons.
 
We model the photoheating during \HeII\ reionization with a multi-zone model that distinguishes the heating from short mean free path photons that nearly instantaneously ionize the medium surrounding a quasar and the heating from long mean free path photons that uniformly heat the rest of the IGM.

\subsubsection{Short mean free path photons}

On average, the short mean free path \HeII\ -ionizing photons that contribute to the formation of \HeIII\ bubbles instantaneously heat the newly-ionized gas by,
\begin{equation}
\Delta Q^{\rm inst}=n_{\rm HeII}\left(\int_{h\nu_{\rm HeII}}^{E_{\rm max}}dE\frac{J_E}{E}\right)^{-1}\int_{h\nu_{\rm HeII}}^{E_{\rm max}}dE(E-h\nu_{\rm HeII})\frac{J_E}{E}, \label{eqn:Qinst}
\end{equation}
where $J_E$ is the average specific intensity that ionized the gas, $h\nu_{\rm HeII}$ is the ionization potential of \HeII\ (4 Ryd/54.4 eV), and $E_{\rm max}$ is the threshold photon energy that distinguishes photons that contribute to a \HeIII\ bubble (short mean free path photons) from photons that free stream and uniformly heat the IGM (long mean free path photons). We approximate that the heat injection from these \HeIII\ bubble-forming photons increases the temperature of the gas around a quasar by $\Delta T_{\rm inst} = 2/ (3 k_b n_{\rm tot})\Delta Q^{\rm inst}$ immediately once the quasar turns on and forms a \HeIII\ bubble. While \HeII\ reionization is underway, the regions that have been exposed to these short mean free path photons are considered fully reionized and no longer experience non-equilibrium photoheating. 
   
\subsubsection{Long mean free path photons}    

\HeII\ -ionizing photons with energies greater than $E_{\rm max}$ do not collectively contribute to ionization fronts during \HeII\ reionization, instead, they free stream for several tens or hundreds of Myr before absorption. Their impact on the global ionization state of the gas is negligible, but the photoionizations from the rare absorptions of these photons heat the IGM approximately homogeneously with a rate of

\begin{equation}
\frac{dQ^{\rm long MFP}_{\rm HeII}}{dt}(z)=n_{\rm HeII}(z)\int_{E_{\rm max}}^{\infty}\frac{dE}{E}(E-E_{\rm HeII})J_E(z)\sigma_{\rm HeII}(E).
\label{eqn:hardbackground}
\end{equation}
The specific intensity of these long mean free path photons at a redshift of $z_0$ is
\begin{equation}
J_E(z_0)=\frac{c}{4\pi}\int_{z_0}^{\infty}dz\left|\frac{dt}{dz}\right|\frac{(1+z_0)^3}{(1+z)^3}\epsilon_E(z)e^{-\tau_{\rm HeII}(z_0,z,E)},
\label{eqn:JE}
\end{equation}
where the optical depth from photons emitted at redshift $z_{\rm em}$ to redshift $z$ is defined by
\begin{align}
 & \tau_{\rm HeII}(z_{\rm em},z,E)  = \nonumber \\  &  \int_{z_0}^{z_{\rm em}}\frac{c \, dz}{H(z)(1+z)}\sigma_{\rm HeII}\left(E\frac{1+z}{1+z_0}\right)\bar{n}_{\rm HeII}(z).
\label{eqn:tau}
\end{align}
In Equation~\ref{eqn:JE}, we approximate the specific ionizing emissivity to hold the shape of a power-law, such that $\epsilon_E=AE^{-\alpha_{\rm QSO}}$, where $\alpha_{\rm QSO}$ is the quasar spectral index (see \S~\ref{sec:qso}). The normalization factor, $A$, is determined by setting the ionizing emissivity equal to the total number of ionizations and recombinations, such that,
\begin{equation}
\int_{E_0}^{\infty}dE\epsilon_E= \bar n_{\rm He}\left(\frac{d\bar{x}_{\rm HeIII}}{dt}+C_{\rm HeIII}\alpha_B\bar{x}_{\rm HeIII} \bar n_e\right),
\label{eqn:normalization}
\end{equation}
where $\bar{x}_{\rm HeIII}$ is the model-dependent mean ionization fraction of \HeIII, and $C_{\rm HeIII}$ the clumping factor (see \S~\ref{sec:C}).

\subsubsection{The patchiness of \HeII\ reionization}
The multi-zone heating based on the two types of photoheating described in the previous subsections illustrates how we implement the thermal impact of \HeII\ reionization. Short mean free path photons form and heat \HeIII\ bubbles, while long mean free path photons uniformly heat all intergalactic gas that is not yet part of a \HeIII\ bubble. Gas particles will be reionized at different times-- \HeIII\ bubbles form around ionizing sources until the bubbles overlap completely and the whole volume ionized. As soon as a gas particle becomes part of a \HeIII\ bubble, photoheating from the long mean free path photon background shuts off and this region experiences the instantaneous heating from the reionization of \HeII\ . This approach of heating and cooling local regions individually avoids artificial cooling rates-- a consequence of using a UV background model to uniformly heat the IGM \cite[e.g.][]{haardt12, puchwein19, fg19}-- as recombination and collisional cooling rates are dependent on the local temperature. 

The gas in regions that are reionized last is exposed to the heating from the long mean free path photons longest and thus is heated more extremely than earlier-reionized regions. \HeII\ reionization ends with significant spatial fluctuations in the temperature of the intergalactic medium. Thus, the duration of \HeII\ reionization will affect the magnitude of these temperature fluctuations.

\subsection{Model implementation}

The morphology of \HeII\ reionization in the radiative transfer simulations of \citet{mcquinn09} suggest that the emergence of \HeIII\ bubbles happens rapidly and that these bubbles occur in a tight distribution of sizes. Our models emulate this process as follows:
\begin{enumerate}
\item{ In the simulation, we identify halos of $M_{\rm halo}\geq10^{12}$M$_{\odot}$\footnote{The mass range for potential quasar host halos is selectable by the user in our model.} and select one of these halos at random to host a quasar.}
\item{ We create a \HeIII\ bubble centered around the quasar when the quasar ``turns on''. The short-mean-free-path heating described in the previous section is applied to gas within the bubble--flash heating the region. Each ionized particle is marked so that particles are only ionized once.}
\item{We turn on additional quasars and form resulting bubbles in the box to match an ionization fraction that is pre-computed at each timestep in the simulation. These ionization fractions are either linear with redshift or are self-consistently calculated using Equation~\ref{eqn:normalization} and a quasar emissivity fitting function (see \S~\ref{sec:evHeII}). Regions that have not been reionized are exposed to the long-mean-free-path photon background.}
\item{ The \HeIII\ bubbles eventually overlap and fill the entire volume. Once a particle is ionized, it stays ionized indefinitely regardless of whether it moves outside the spatial extent of the \HeIII\ bubble at formation. Once the desired ionization fraction exceeds $0.95$, all remaining gas in the simulation volume is ionized.}
\end{enumerate}

Because the uniform long mean free path heating is dependent only on the parameters of \HeII\ reionization selected prior to running the simulation, this heating is calculated ahead of time in tandem with the \HeII\ reionization evolution. These radiative transfer calculations are run in a separate (public) code that takes the user-selected parameters of \HeII\ reionization and produces an input file for the simulation.

\subsection{Free parameters}
We aim to capture the extent of the parameter space of \HeII\ reionization in these models. The following free parameters encapsulate the uncertainties in the sources of \HeII\ -ionizing photons and the evolution of the thermal state of the IGM. 

\subsubsection{Quasar spectral index}
\label{sec:qso}
The emissivity of quasars at 4 Ryd is uncertain. This owes to the difficult nature of observing any extragalactic radiation in the extreme ultraviolet and the uncertainty in the quasar SED at these energies. Observations in the nearer UV only reliably estimate the shape of this SED to a couple tens of electron volts \cite[e.g.][]{telfer02}. The heating resulting from \HeII\ reionization is highly dependent on this emissivity. Most estimates of this emissivity model quasar SEDs as single power-laws fixed at 912$\AA$.

Quasar stacks in the UV are used to constrain the power-law index of the spectral energy distribution, $\alpha_{\rm QSO}$. The analyses of these stacks find varying values of $\alpha_{\rm QSO}$, between $\sim 1.0$ and $2.0$ \citep{telfer02,shull12,stevans14,lusso15}. Uncertainty in the \HI\ absorption from the IGM and the subsequent correction accounts for much of this discrepancy. Additionally, a pivot wavelength must be selected to construct these composite spectra and the spectral coverage in the rest frame of the quasar varies by redshift-- another source of the differences between measurements of $\alpha_{\rm QSO}$.

Unlike \HI\ reionization, collisional cooling is not effective in cooling the \HeII\ ionization fronts, and thus the value of $\alpha_{\rm QSO}$ drives the magnitude of heating from \HeII\ reionization. Because $\alpha_{\rm QSO}$ is the largest source of uncertainty for the temperature of a \HeIII\ bubble immediately after it has undergone reionization, the value of $\alpha_{\rm QSO}$ is crucial in determining the magnitude of peak of the temperature at $z\sim3$. 

\subsubsection{Evolution and duration of HeII reionization}
\label{sec:evHeII}

\begin{figure}
\begin{center}
\resizebox{9.0cm}{!}{\includegraphics{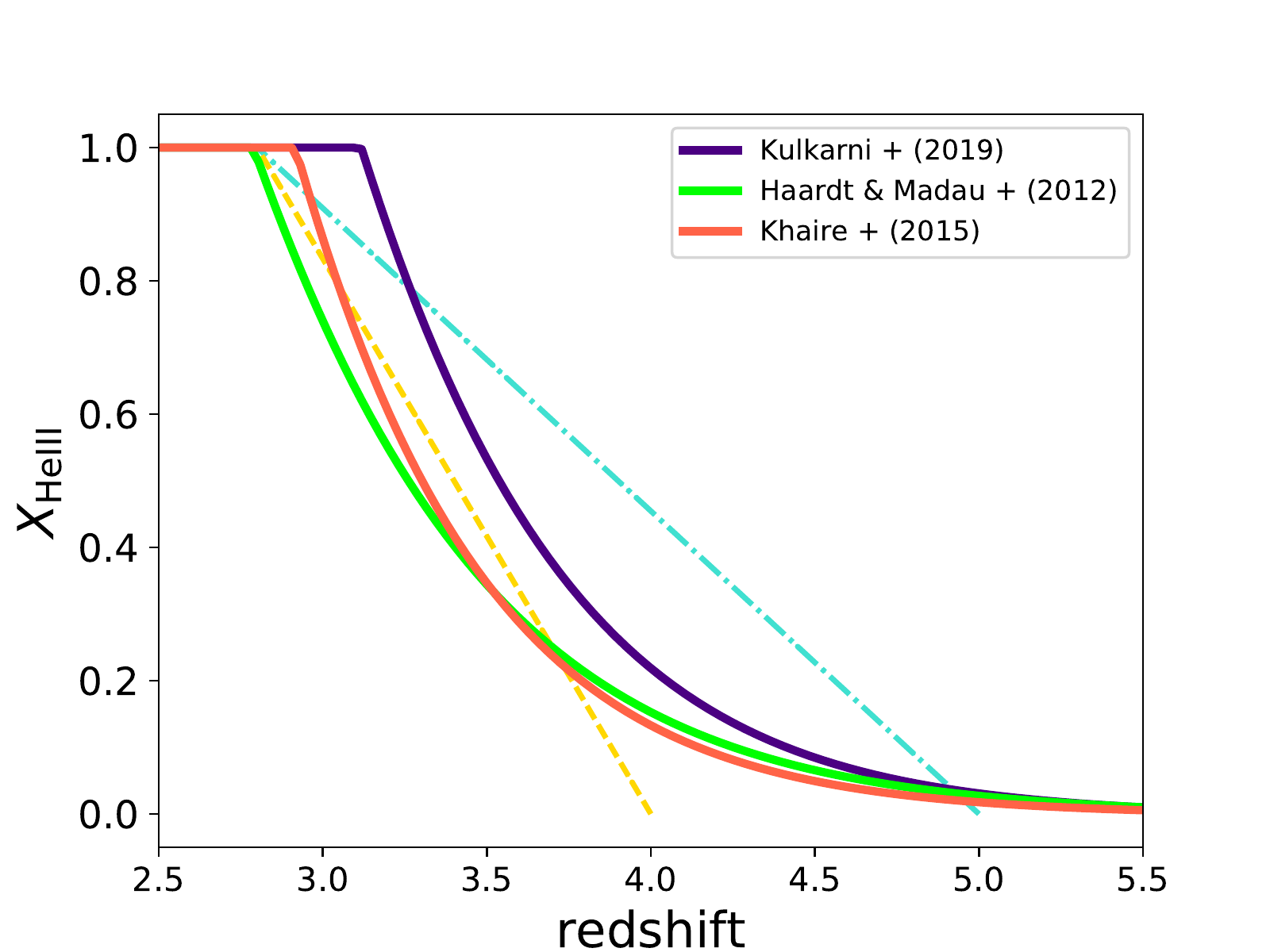}}\\
\end{center}
\caption{A sampling of \HeII\ reionization histories available within our flexible \HeII\ reionization model. The solid curves show the \HeII\ reionization histories based on the quasar emissivity functions from \citet{kulkarni19}, \citet{khaire15}, and \citet{haardt12}. The dashed lines show our simple linear reionization histories with $2.8 <z_{\rm rei} < 4$ and $2.8 <z_{\rm reion,HeII} < 5$.
\label{fig:XHeIII}}
\end{figure}

While the \HeII\ Lyman $\alpha$ forest provides a rough estimate in redshift for the finish of \HeII\ reionization, the duration and evolution is still uncertain. Large error bars on temperature measurements at high redshift and quasar luminosity functions do not reveal an obvious onset, though a very early start to \HeII\ reionization would be hard to justify in the face of most of the low temperature measurements at higher redshifts ($z>4.5$).

We model the evolution of \HeII\ reionization in one of two ways-- (1) an ionization fraction that evolves linearly with redshift with the starting and ending redshifts left as free parameters; (2) a reionization evolution based on a given fitting formula for the quasar emissivity (e.g. \citealt{haardt12,khaire15,kulkarni19}) and a clumping factor (see \S~\ref{sec:C}). We self-consistently calculate the emissivity-based reionization histories using Equation~\ref{eqn:normalization}-- examples of which are are shown in as the solid curves in Figure~\ref{fig:XHeIII}. The dashed lines in Figure~\ref{fig:XHeIII} show sample linear reionization histories with $2.8 <z_{\rm reion,HeII} < 4$ and $2.8 <z_{\rm reion,HeII} < 5$. Though our models that use a reionization history based on a given fitting formula for the quasar emissivity are more physically motivated, our fiducial model uses a linear history with $2.8 <z_{\rm reion,HeII} < 4$. We use this simple model because the ionization evolution results in a thermal history most closely in agreement with the thermal history we find using our uniform UVB model; thus we can better isolate the effect of the inhomogeneity of \HeII\ reionization.

\begin{figure*}
\begin{center}
\resizebox{15.0cm}{!}{\includegraphics{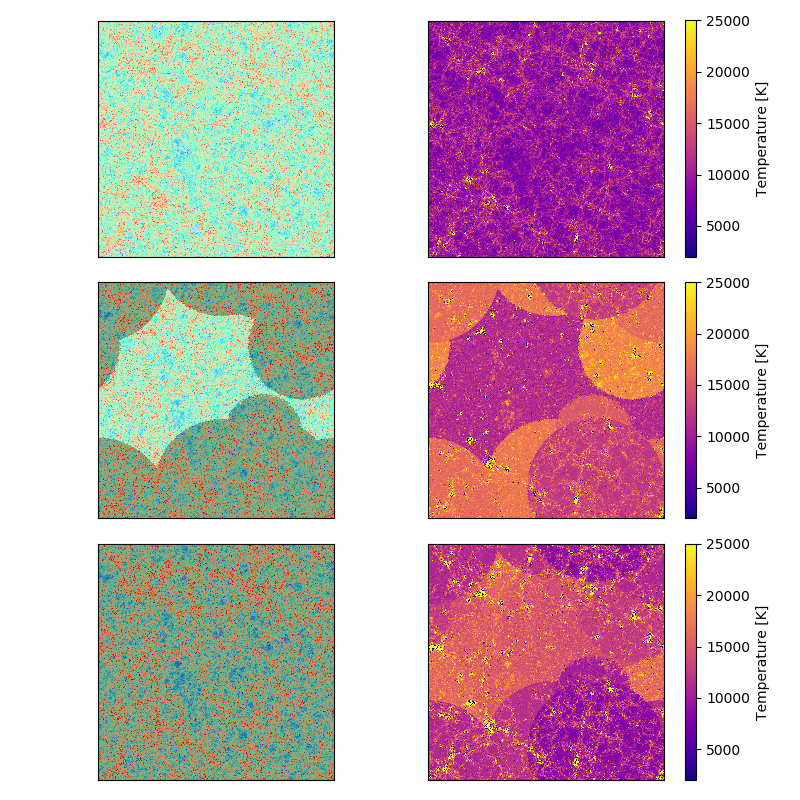}}\\
\end{center}
\caption{Snapshots from a simulation with our fiducial \HeII\ reionization model at $z=4$ (top panel), $z=3.25$ (middle panel), and $z=2.5$ (bottom panel). The left panels show slices of the density distribution, where the \HeII\ -ionized regions are shown as shaded. The right panels shows slices of the temperature of the IGM. This fiducial model sets $\alpha_{\rm QSO} = 1.7$, $C_{\rm HeIII } = 3$, $E_{\rm max} = 150$ eV, and $z_{\rm reion,HeII}=2.8-4$.
\label{fig:snaps}}
\end{figure*}

\subsubsection{\HeIII\ bubble size}

The size distribution of the \HeIII\ bubbles will be dependent on the properties of their sourcing quasars-- most importantly their luminosity and lifetime. In the radiative transfer simulations of \citet{mcquinn09}, the quasars are modeled with two different methods: (1) quasars have a constant lifetime and their luminosity is dependent on the host halo mass, and (2) the quasar lifetime is dependent on the luminosity. The distribution of resulting \HeIII\ bubble sizes is centered around 20 comoving Mpc (for a quasar lifetime of 40 Myr) and 35 comoving Mpc respectively. The latter method has a much tighter distribution because the relation between quasar luminosity and lifetime results in brighter quasars having shorter lifetimes. Though these quasar models are extremely simple, they give a reasonable first order estimation. We leave both the mean bubble size and variance in the distribution as free parameters. The default bubble size is $30$ Mpc/$h$ with $0$ variance. 

\subsubsection{Threshold energy of \HeIII\ bubble photons}

Equation~\ref{eqn:mfp} estimates the photon energy which would give a mean free path longer than the size of the \HeIII\ bubbles. Though \citet{uptonsanderbeck16} found that the mean thermal history was highly insensitive to the value of this threshold photon energy that separates the short mean free path photons that contribute to \HeIII\ bubbles from the uniform long mean free path photon background, it is dependent on the size of the mean \HeIII\ bubble size. Because the \HeIII\ bubbles are likely on the order of tens of Mpc, the value of this threshold energy, $E_{\rm max}$, should fall roughly between 100-200 eV. 

\subsubsection{Clumping factor}
\label{sec:C}
The clumping factor, $C_{\rm HeIII }$, describes an excess of \HeIII\ recombinations over a homogeneous Universe with $T=10^4$ K, defined by
\begin{equation}
C_{\rm HeIII} = \frac{\langle \alpha_{\rm B} n_{\rm HeIII} n_{e}\rangle}{\langle \alpha_{\rm B}\rangle \langle n_{\rm HeIII}\rangle\langle n_{e}\rangle}
\end{equation}
where $\alpha_{\rm B}$ is the Case B recombination rate for \HeIII\ , and $n_{e}$ and $n_{\rm HeIII}$ are the number densities of electrons and \HeIII\ respectively. The Case B recombination rate is used as an approximation, as during \HeII\ reionization, recombinations to the ground state will result in another ionizing photon that will be likely be locally reabsorbed by remaining \HeII\ . The clumping factor is found on the right side of Equation~\ref{eqn:normalization}, and modulates the normalization of the emissivity of ionizing sources and heating from the long mean free path photon background due to deviations from homogeneity. $C_{\rm HeIII }$ is typically in the range of $\sim1-5$, near unity at high redshift and increasing with decreasing redshift. During the span of \HeII\ reionization, we have calculated the clumping factor in our simulations to grow between $2\lesssim C_{\rm HeIII }\lesssim 3.5$. Despite this subtle evolution, in these simulations, we use a constant value-- selectable by the user. While the clumping factor would ideally be calculated on-the-fly, the radiative transfer calculations detailed in \S~\ref{sec:thermal} are performed prior to running the simulation to avoid {\it in situ} radiative transfer, and thus it must be selected {\it a priori}.

\section{Simulations and results}
\label{sec:results}

The code described in this paper is integrated into the cosmological simulation code MP-Gadget \cite{yu_feng_2018_1451799}. MP-Gadget is descended from Gadget-2 \cite{Springel:2005}, with heavy modifications for scalability. We recommend using this code with simulations that exceed a box size of $\sim$100 Mpc in order to fully capture the scales of \HeII\ reionization effects. We use a fiducial UV background model from \citet{puchwein19}, modified to have zero \HeII\ photoheating during \HeII\ reionization. The simulations shown in \S~\ref{sec:results} are run with a box size of 100 Mpc/$h$ and 1024$^3$ particles. Cooling, star formation, a black hole accretion model, and feedback from supernovae and black holes are included following \citet{feng16}.

In this section, we present the results of our simulations, demonstrate the impact of \HeII\ reionization on various gas statistics, and compare these results with historical \HeII\ reionization radiative transfer simulations. Figure~\ref{fig:snaps} shows snapshots of our simulation with fiducial parameters at a redshift prior to \HeII\ reionization ($z=4$), during \HeII\ reionization ($z=3.25$), and after \HeII\ reionization has completed ($z=2.5$). The left panels show the density distribution with shaded \HeII\ -ionized regions. The right panel shows the corresponding temperature field. Our fiducial model sets $\alpha_{\rm QSO} = 1.7$, $C_{\rm HeIII } = 3$, $E_{\rm max} = 150$ eV, and $z_{\rm reion,HeII}=2.8-4$.

\subsection{Evolution of the thermal history of the IGM at the mean density}

\begin{figure}
\begin{center}
\resizebox{9.5cm}{!}{\includegraphics{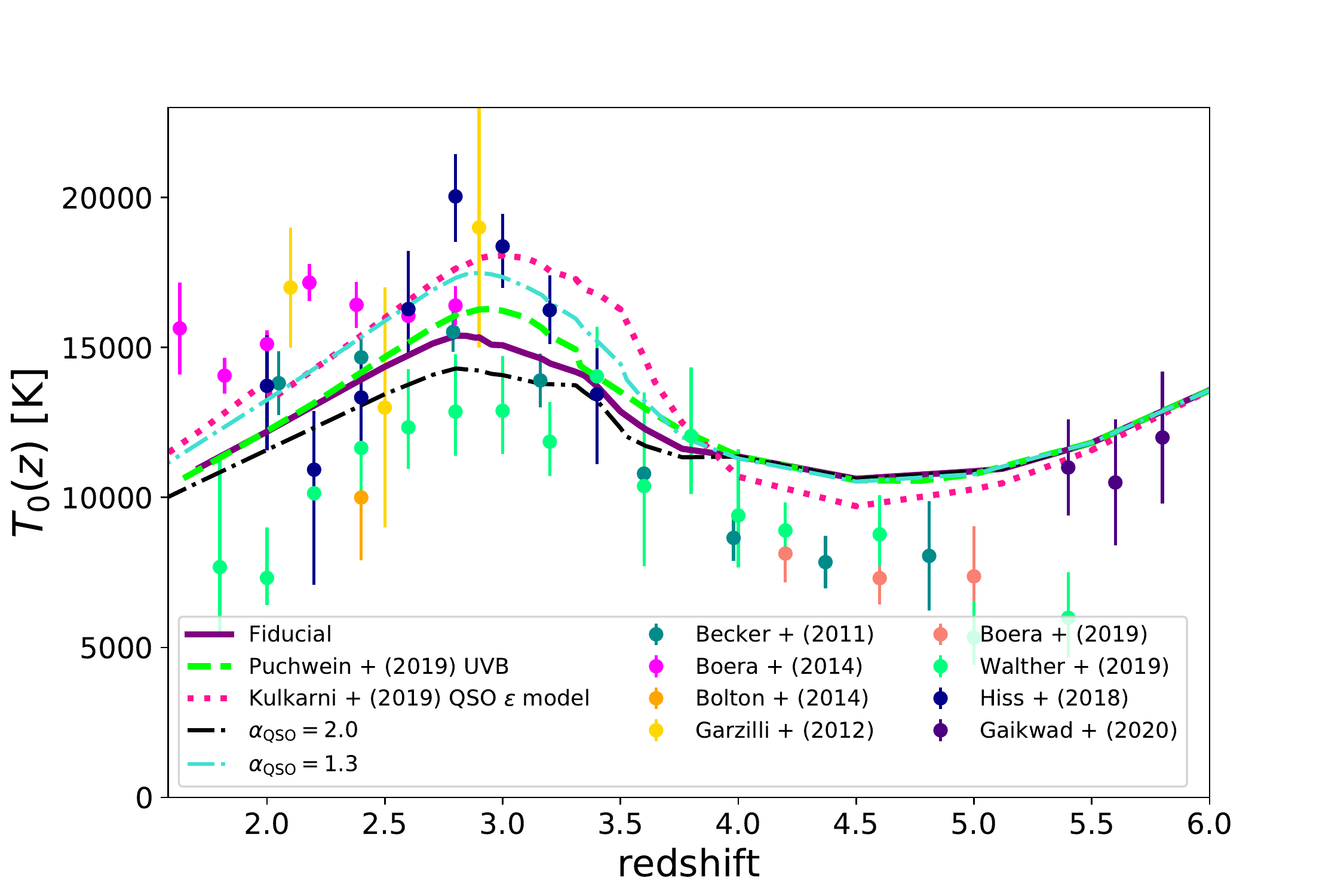}}\\
\end{center}
\caption{Average temperature evolution of the IGM at the cosmic mean density for a selection of \HeII\ reionization models. The solid purple curve shows our fiducial model with $\alpha_{\rm QSO} = 1.7$, $C_{\rm HeIII } = 3$, $E_{\rm max} = 150$ eV, and $z_{\rm reion,HeII}=2.8-4$. The dashed turquoise and black curves vary $\alpha_{\rm QSO}$ to 1.3 and 2.0 respectively. The dotted pink curve shows a thermal history based on the quasar emissivity function of \citet{kulkarni19}. The dashed green curve shows the temperature evolution without an inhomogeneous \HeII\ reionization model, but that recreates the mean thermal history with the Puchwein et al. (2019) UV background model.
\label{fig:t0}}
\end{figure}

The evolution of the thermal state of intergalactic gas has been studied extensively over the past couple decades in response to the emergence of high-resolution, high signal-to-noise Lyman $\alpha$ forest spectra data. These spectra allowed for measurements of the temperature of the low density IGM through quantification of the small-scale structure in the absorption features. A bump found in the temperature at $z\sim3$ hinted at additional photoheating from \HeII\ reionization, though variations in temperature measurements amongst different studies have disputed the amplitude and timing of this peak \citep{schaye00, lidz10,becker11,garzilli12,boera14,bolton14,hiss18,walther19,boera19,gaikwad20}. 

Recent UV background models \cite[e.g.][]{haardt12, puchwein19} are able to emulate the evolution of the temperature of the IGM at the mean cosmic density, $T_0(z)$, which includes a bump in temperature from \HeII\ reionization. However, modeling the impact of \HeII\ reionization with a single, spatially uniform UVB model will result in heating and cooling rates that differ from those found in an inhomogeneous reionization. Not only do these models miss the impulsive heating during the creation of \HeIII\ bubbles, but many of the cooling processes of intergalactic gas are dependent on temperature, such as from recombinations and collisional cooling. 

Figure~\ref{fig:t0} shows the temperature evolution at the mean cosmic density for a selection of our \HeII\ reionization models and a single case with only the \citet{puchwein19} UV background model. The solid purple curve shows our fiducial model with a linear reionization history that spans $2.8 < z_{\rm reion,HeII} <4.0$, $\alpha_{\rm QSO} = 1.7$, $C_{\rm HeIII } = 3$, and $E_{\rm max} = 150$ eV. The dashed turquoise curve and dashed black curve show this model, but with $\alpha_{\rm QSO}$ varied to 1.3 and 2.0 respectively. The dotted pink curve shows our fiducial model, but with a \HeII\ reionization history based on the quasar emissivity function of \citet{kulkarni19}. Finally, the dashed green curve shows the temperature evolution resulting from the \citet{puchwein19} UV background model. The various error bars show temperature measurements from the previous decade, namely \citet{lidz10}, \citet{becker11}, \citet{garzilli12}, \citet{boera14}, \citet{bolton14}, \citet{hiss18}, \citet{boera19}, \citet{walther19}, and \citet{gaikwad20}. While there is only marginal agreement amongst the temperature measurements, the general shape and peak at $z\approx3$ is in agreement with the parameter space allowed in our simulations. 

\subsection{Temperature-density relation}
The temperature-density relation is traditionally modeled by a power-law of the form, 
\begin{equation}
T = T_0 \Delta^{\gamma - 1}
\end{equation}
where $T_0$ is the temperature of the IGM at the mean cosmic density, $\Delta$ is the fractional gas overdensity, and $\gamma - 1$ is the power-law index shaped by the heating and cooling processes of the gas. This parameterization holds generally true after \HeII\ reionization and possibly after hydrogen reionization\footnote{If hydrogen reionization ends lower than $z\approx 6$ as recent findings may suggest, the gas may have residual temperature fluctuations from reionization and will not conform to a tight power law temperature-density relation before the onset of \HeII\ reionization.}, settling approximately into a single power-law with $\gamma$ typically holding a value of $\approx 1.6$ \citep{gnedin97,mcquinn16}. Once \HeII\ reionization begins, the regions undergoing reionization are flash heated by thousands of degrees, while the newly \HeIII\ -ionzed regions begin to cool. Regions where \HeII\ has not yet been ionized are exposed to the long mean free path photons that gradually heat the gas. The spatially-varying heating and cooling introduces scatter in the temperature-density relation that peaks around the end of \HeII\ reionization.

\begin{figure*}
\begin{center}
\includegraphics[width=0.9\textwidth]{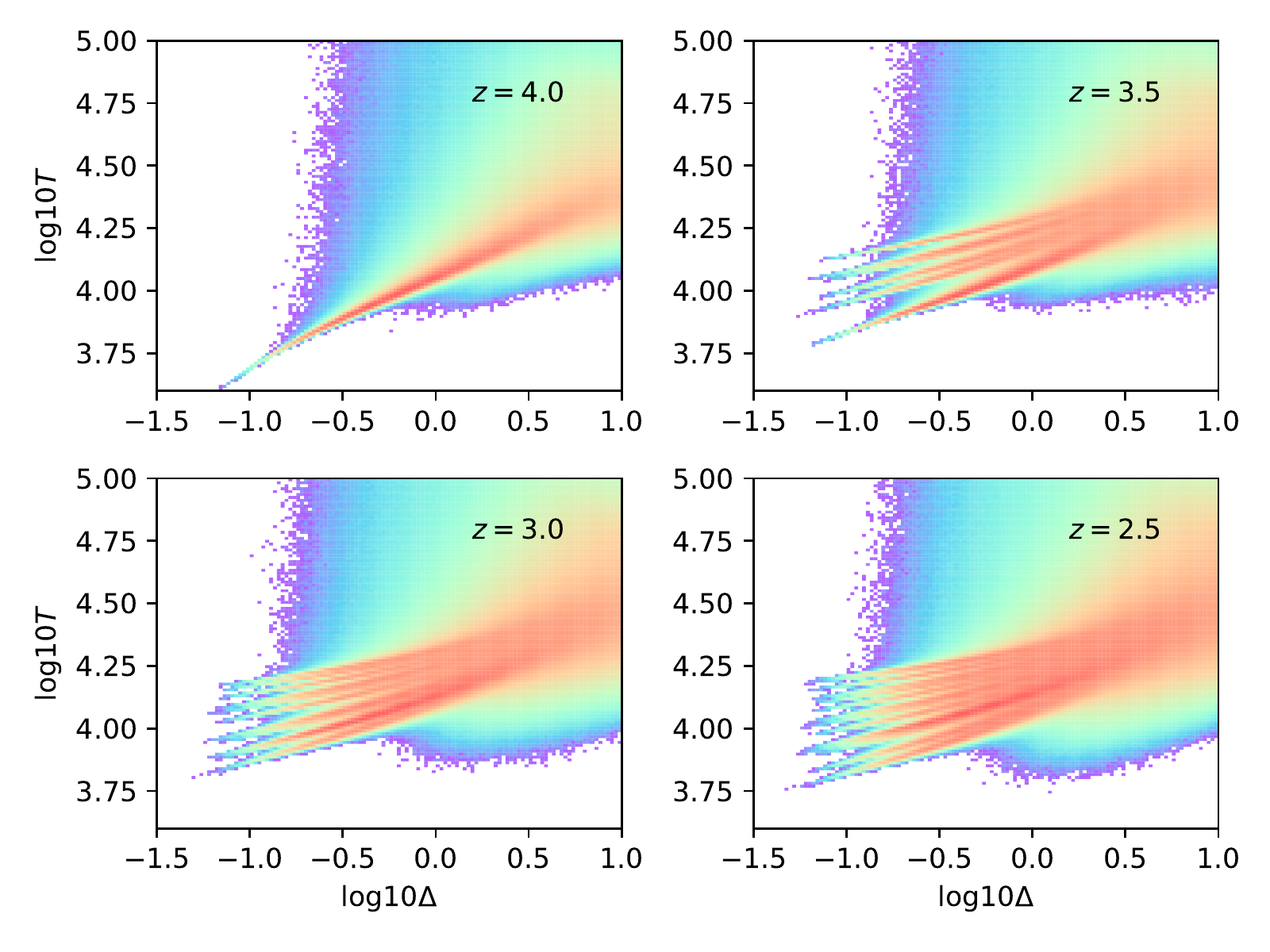}
\end{center}
\caption{Temperature-density relation of the low density IGM with our fiducial \HeII\ reionization model at $z=4.0$, $3.5$, $3.0$, and $2.5$. In this model, \HeII\ reionization begins at $z=4$ and ends at $z = 2.8$. As different regions become \HeII\ -ionized at different redshifts, the temperature-density relation branches to reflect the independently evolving \HeIII\ bubbles. The colorscale represents a logarithmic number of particles within a two-dimensional bin.
\label{fig:tdelt}}
\end{figure*}

The evolution of the temperature-density relation of the IGM is shown in Figure~\ref{fig:tdelt} for our fiducial \HeII\ reionization model. The dispersion seen in the bottom panels of Figure~\ref{fig:tdelt} is marked by several distinct power-laws that represent each independently-evolving \HeIII\ bubble.\footnote{Realistically, \HeIII\ bubbles are not instantaneously ionized, so dispersion in the temperature-density should be smoother than our simulations suggest.} The scatter in the temperature-density relation will peak at the conclusion of \HeII\ reionization. Following the end of the photoheating associated with reionization, the temperature-density relation will graudually settle back down into a single power-law. 

Many of the temperature measurements aim to simultaneously constrain the temperature at the mean cosmic density and $\gamma$, with the assumption that the temperature-density relation follows a single power-law relation, but this does not reflect the nature of the independently-evolving regions that are ionized at different epochs. However, the dispersion in the temperature-density relation is highest at the lowest densities, and thus the observability of these fluctuations is diminished \citep{mcquinn11}.

\subsection{The Lyman $\alpha$ forest, flux PDF, and flux power spectrum}

The flux probability distribution function (PDF) and the one-dimensional flux power spectrum of the Lyman $\alpha$ forest are the most frequently used statistics of this observable. On one end (small $k$), it is sensitive to the large-scale density distribution of matter and at large $k$, it is more sensitive to astrophysical processes such as ionizing backgrounds and thermal fluctuations in intergalactic gas. Because the Lyman $\alpha$ forest flux power spectrum quantifies fluctuations at both cosmological and astrophysical scales, it is a potential tool to constrain cosmological parameters. 

\begin{figure}
\begin{center}
\resizebox{9.0cm}{!}{\includegraphics{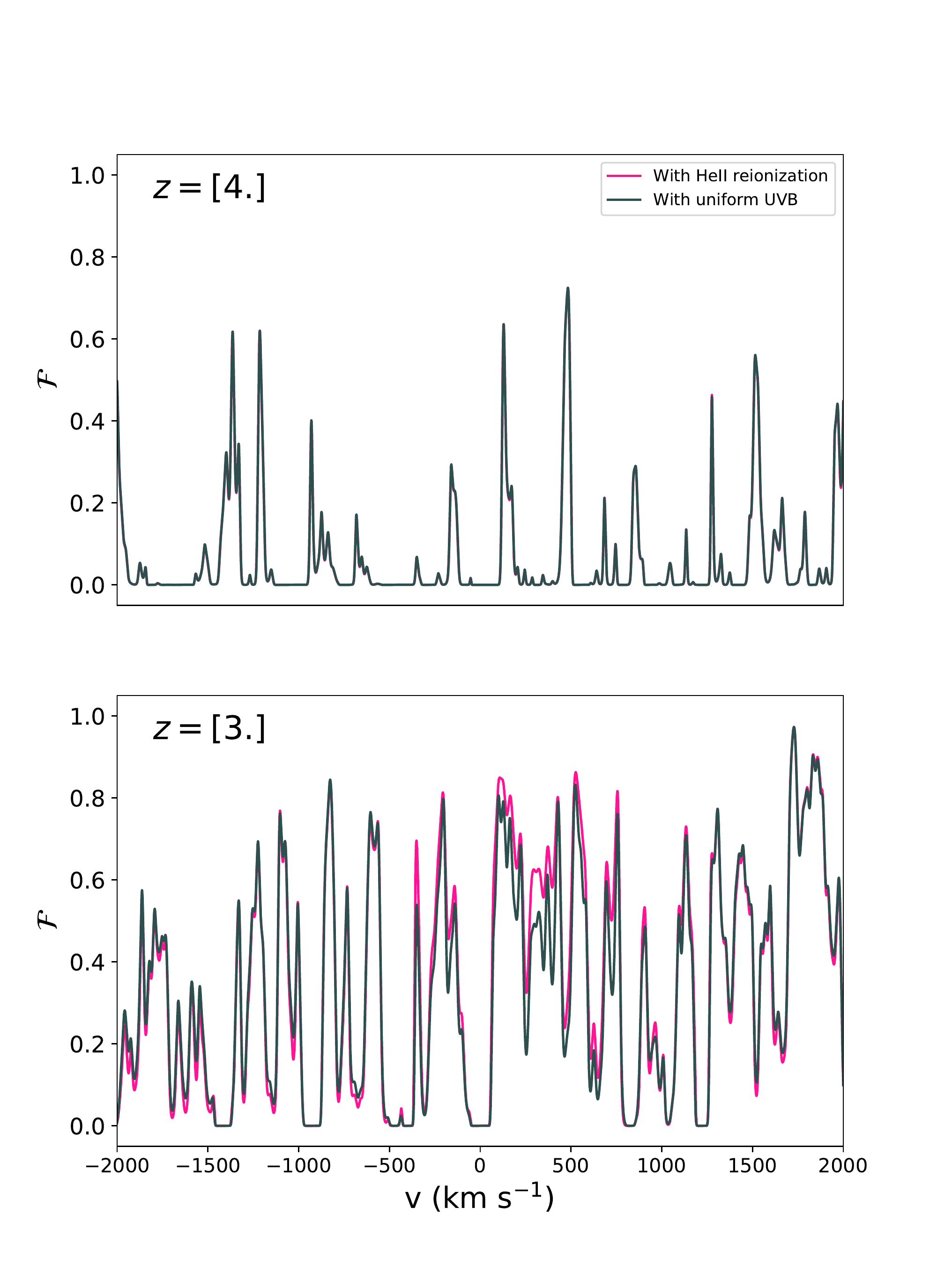}}\\
\end{center}
\caption{The simulated \HI\ Lyman $\alpha$ forest of our \HeII\ reionization model in pink and a model with a uniform UVB in grey. The top panel shows the forest at the onset of \HeII\ reionization in our model ($z=4$), and the bottom panel shows a redshift towards its completion ($z=3$).
\label{fig:lya}}
\end{figure}

\begin{figure*}
\begin{center}
\includegraphics[width=0.8\textwidth]{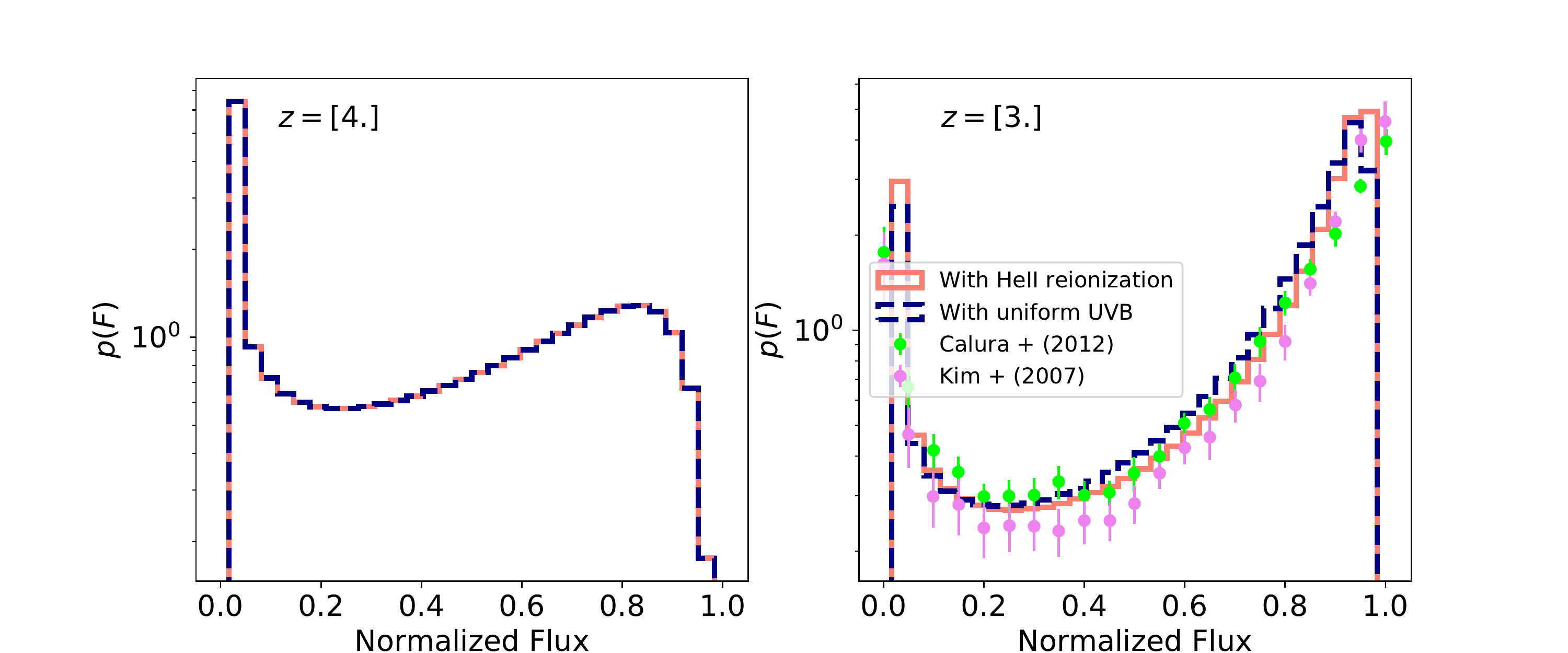}\\
\end{center}
\caption{Flux probability distribution function from a simulation with inhomogeneous \HeII\ reionization model (solid) and from a simulation with a uniform UVB (dashed). The left panel shows a redshift at the onset of \HeII\ reionization ($z=4$) and the right panel shows a redshift just before its completion ($z=3$). The error bars show measurements from \citet{kim07} and \citet{calura12}.
\label{fig:pdf}}
\end{figure*}

\begin{figure*}
\begin{center}
\resizebox{13.5cm}{!}{\includegraphics{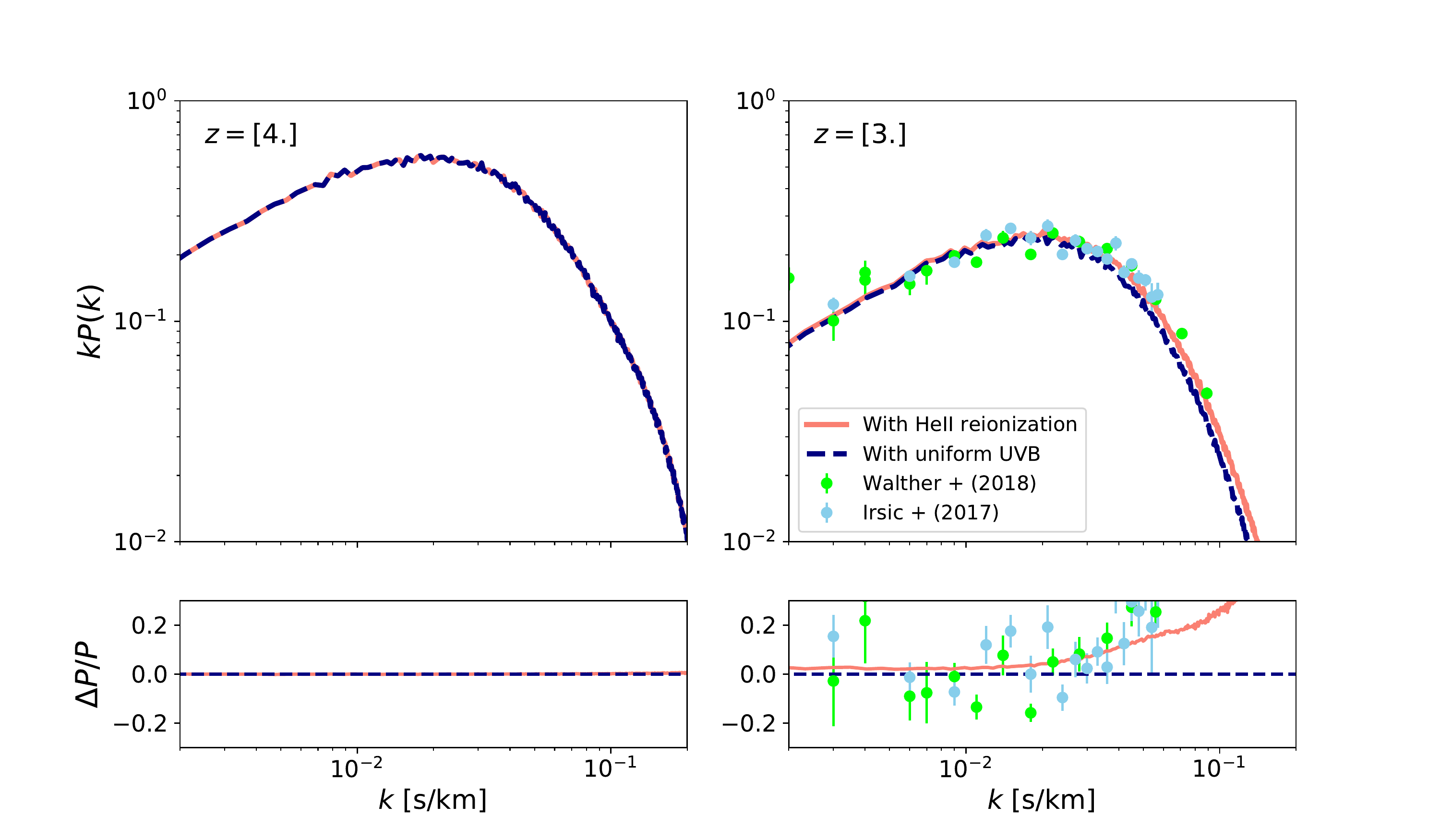}}\\
\end{center}
\caption{One dimensional flux power spectrum from a simulation with inhomogeneous \HeII\ reionization model (solid) and from a simulation with a uniform UVB (dashed). The lower panels show the fractional change between these two models. The left panel shows a redshift at the onset of \HeII\ reionization ($z=4$) and the right panel shows a redshift just before its completion ($z=3$). The error bars show measurements from \citet{irsic17} and \citet{walther18}. 
\label{fig:ps}}
\end{figure*}

In Figures~\ref{fig:lya}, ~\ref{fig:pdf}, and ~\ref{fig:ps}, we compare (1) a simulation that has our \HeII\ reionization model turned off, but with the \citet{puchwein19} UVB model uniformly heating the IGM during the redshifts of \HeII\ reionization and (2) a simulation with our \HeII\ reionization model that roughly matches the temperature evolution at mean cosmic density of the former. Recalling Figure~\ref{fig:t0}, the closest match to the \citet{puchwein19} UVB model is our fiducial model-- thus this the \HeII\ reionization model we use for the remainder of the figures, unless accompanied by additional models. Because we are approximately controlling for the temperature evolution of intergalactic gas at the mean density during \HeII\ reionization, the following comparisons highlight the effect of patchy reionization on the Lyman $\alpha$ forest.  Figure~\ref{fig:lya} shows simulated Lyman $\alpha$ forest spectra with and without inhomogeneous \HeII\ reionization. Because the individual lines in the Lyman $\alpha$ forest are sensitive to the local thermal state of intergalactic gas, spectra from models with patchy \HeII\ reionization show more dispersion in the small scale structure of the absorption features. This can be more clearly represented in the following statistics of the forest. 

\begin{figure}
\begin{center}
\resizebox{9.5cm}{!}{\includegraphics{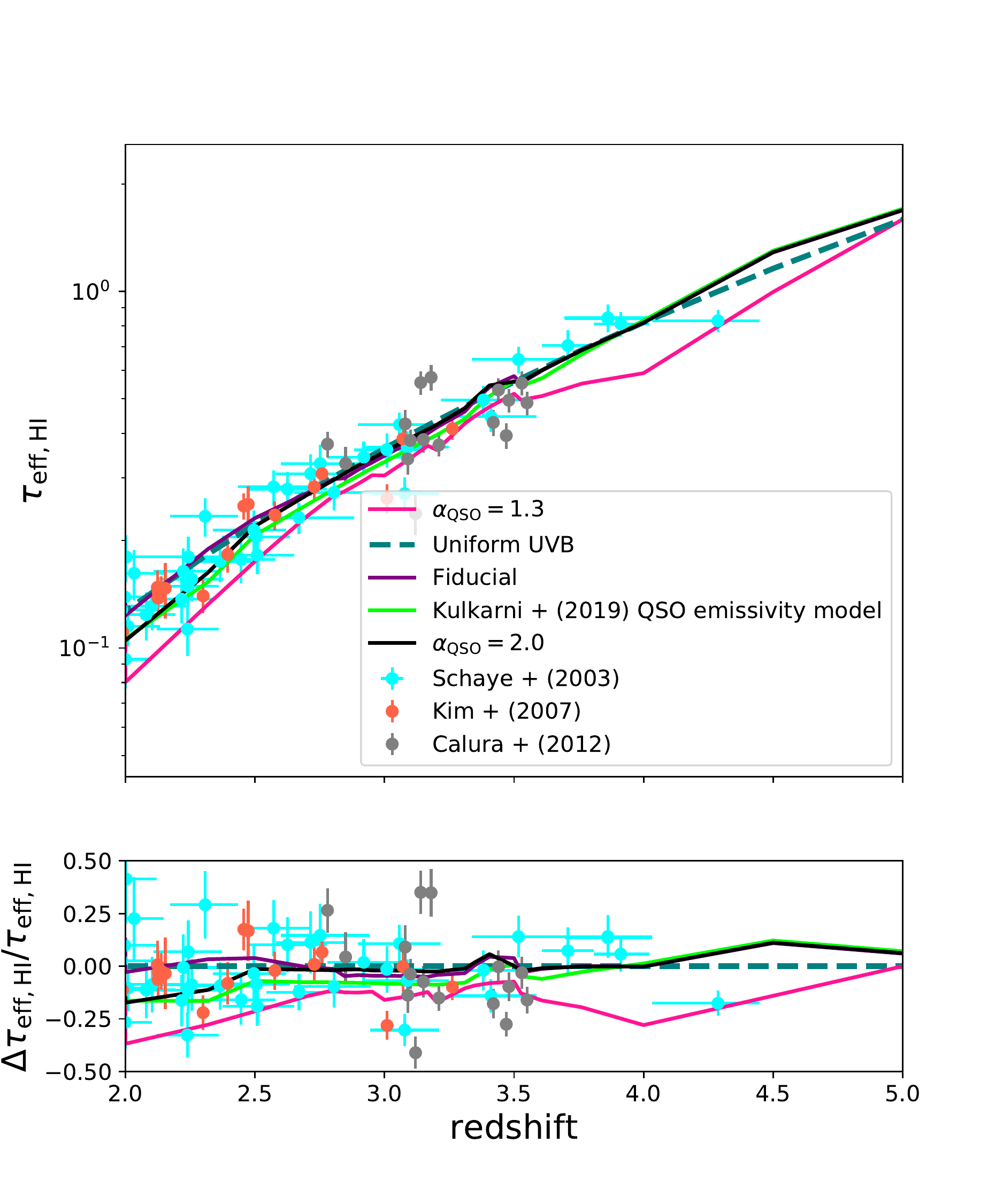}}\\
\end{center}
\caption{ The evolution of the effective optical depth of the \HI\ Lyman $\alpha$ forest for three of our \HeII\ reionization models compared with a model with a uniform UVB (dashed, teal curve). The solid purple curve shows our model with $\alpha_{\rm QSO} = 1.7$, $C_{\rm HeIII } = 3$, $E_{\rm max} = 150$ eV, and $z_{\rm reion,HeII}=2.8-4$. The pink and black curves shows the same model, but with $\alpha_{\rm QSO}=1.3$ and $\alpha_{\rm QSO}=2.0$, respectively. The light green curve shows our model based on the quasar emissivity function of \citet{kulkarni19}. The error bars show measurements from \citet{schaye03}, \citet{kim07}, and \citet{calura12}. The bottom panel shows the relative differences between our four \HeII\ reionization models and the uniform UVB model. 
\label{fig:taueff}}
\end{figure}

We compare the evolution of the effective optical depth of the simulated \HI\ Lyman $\alpha$ forest between models with and without inhomogeneous \HeII\ reionization in Figure~\ref{fig:taueff}. The dashed teal curve shows our uniform \citet{puchwein19} UVB model-- scaled to the effective optical depth fitting function of \citet{kim07}. The solid curves represent our fiducial model (purple), our model with $\alpha_{\rm QSO}=1.3$ (pink), with $\alpha_{\rm QSO}=2.0$ (black), and our model with an ionization fraction based on the quasar emissivity function of \citet{kulkarni19} (green). These curves have all been scaled by the same factor as our uniform UVB model. The effective optical depth varies up to tens of percent, with our $\alpha_{\rm QSO} = 1.3$ model showing the largest discrepancy due to the largest variance in IGM temperature. 

From our simulated \HI\ Lyman $\alpha$ forest spectra, we calculate the flux PDF for our two simulations shown in Figure~\ref{fig:lya}. Because the patchiness of reionization will result in multiple temperatures at the same gas density, the flux PDF will reflect more bimodality after \HeII\ reionization is underway. The left panel of Figure~\ref{fig:pdf} shows the flux PDF for a redshift at the onset of \HeII\ reionization ($z=4$) and towards the end of reionization in the right panel ($z=3$). The solid curve shows our model with inhomogeneous reionization and the dashed curve shows the effect of a homogeneous UVB. Indeed, in the midst of \HeII\ reionization, we find a more bimodal distribution in comparison to our simulation without inhomogeneous \HeII\ reionization. We compare these cases with observational measurements of the flux PDF from \citet{kim07} and \citet{calura12} at $z\approx3$. While these measurements show broad agreement with both simulations, and may slightly favor our inhomogeneous \HeII\ reionization, the uncertainty in the measurements is nearly equal to, or greater than the difference between our model and the uniform UVB case.

From these same simulations, we show the one dimensional flux power spectrum in Figure~\ref{fig:ps}. The left panel presents the flux power spectrum for a redshift at the onset of \HeII\ reionization ($z=4$) and the right panel shows how the two models deviate after \HeII\ reionization is well underway ($z=3$). The bottom panels show the fractional difference between our model with inhomogeneous reionization and a homogeneous UVB. These power spectra are calculated from two thousand sightlines that have been collectively scaled to the observed mean flux of \citet{kim07}. The inclusion of \HeII\ reionization changes the amplitude of the flux power spectrum up to the several percent level: a few percent at scales approximately greater than $k \approx 10^{-1.5}$ s/km, and growing to tens of percent at scales $\gtrsim 10^{-1.5}$ s/km. At small scales (large $k$), the changes in power are driven by both the thermal state of the gas and thermal history-dependent pressure smoothing. At large scales (small $k$), the power is modulated due to the large spatial fluctuations in temperature from our inhomogeneous reionization model. Though we approximately controlled for the temperature at the mean cosmic density between these two simulations, we acknowledge that some of the deviation may owe to small differences in the evolution of $T_0$. We compare the discrepancy between these models to the power spectrum measurements of \citet{walther18} and \citet{irsic17}, and find broad agreement with both of our simulations at small $k$, with our inhomogeneous \HeII\ reionization model favored at larger $k$. Despite this, due to the tens-of-percent-level error
bars and differences between the separate measurements, neither model is decidedly ruled out.\\

\citet{mcquinn11} predicted that the effect of temperature fluctuations from \HeII\ reionization on these one-dimensional statistics would be exceedingly small. However, their model does not include the dynamical response of intergalactic gas to the thermal impact of \HeII\ reionization such as self-consistent pressure smoothing of the gas. This pressure smoothing is a natural outcome of our model and potentially contributes to the impact of our \HeII\ reionization model. Pressure smoothing is included in the recent  \HeII\ reionization radiative transfer simulations of \citet{laplante17}, who find a similar effect of inhomogeneous \HeII\ reionization on the one dimensional flux power spectrum to our Figure~\ref{fig:ps}. However, pressure smoothing is predicted to have a sub-dominant effect at small scales (large $k$) to the response of the power spectrum to the thermal state of the gas \citep{gnedin98}. Despite the small effect on the one-dimensional statistics of the forest, precision cosmological searches now push constraints to the percent-level-- benefitting from even small improvements in such statistics.

\subsection{\HeII\ Lyman $\alpha$ forest}

Though more observationally elusive than the \HI\ Lyman $\alpha$ forest, the \HeII\ Lyman $\alpha$ forest is an ideal diagnostic for the timing and duration of \HeII\ reionization. The \HeII\ Lyman $\alpha$ line falls at $303.8\AA$, within the extreme ultraviolet, and thus its forest is only possible to observe above $z>2$ -- at lower redshifts, our own galaxy's high column density systems absorb any \HeII\ Lyman $\alpha$ transmission. Even more limiting, quasar sightlines that intersect systems with $N_{\rm HI} \gtrsim 10^{17}$ cm$^{-2}$ have absorbed \HeII\ Lyman $\alpha$ spectral regions. These limitations leave a small fraction of quasar sightlines that have usable \HeII\ Lyman $\alpha$ forest-- currently only a few dozen show sufficient transmission \citep{worseck11,syphers13,syphers14,zheng15,worseck16,worseck19}. 

\begin{figure}
\begin{center}
\resizebox{9.0cm}{!}{\includegraphics{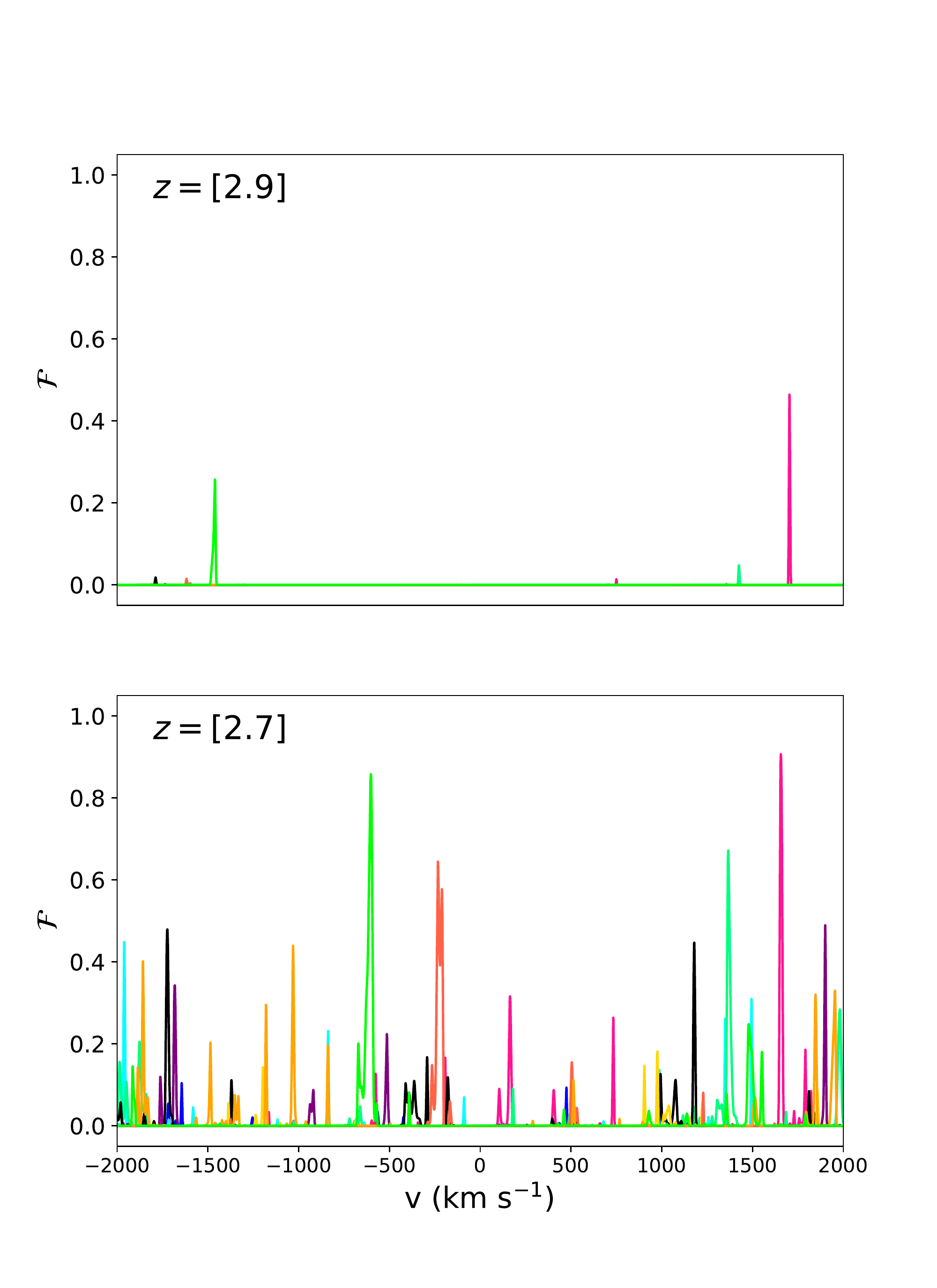}}\\
\end{center}
\caption{Simulated \HeII\ Lyman $\alpha$ forest spectra from our fiducial \HeII reionization. The top panel shows ten different sightlines at $z=2.9$ (before the completion of \HeII\ reionization) and the bottom panel shows ten sightlines from $z = 2.7$ (after the completion of \HeII\ reionization).
\label{fig:heiilya}}
\end{figure}

Because our simulations track the different ionization states of helium and capture the patchiness of the \HeII\ field, we are able to extract mock \HeII\ Lyman $\alpha$ forest spectra. Figure~\ref{fig:heiilya} shows the \HeII\ Lyman $\alpha$ forest from a simulation with our fiducial \HeII\ reionization model at multiple redshifts-- toward the conclusion of \HeII\ reionization ($z=2.9$) and following the completion ($z=2.7$). These spectra are qualitatively consistent with the existing \HeII\ Lyman $\alpha$ forest sightlines, showing a limited number of transmission spikes at redshifts where \HeII\ reionization is still ongoing that likely correspond to very underdense regions that have already undergone \HeII\ reionization. The sightlines shown in Figure~\ref{fig:heiilya} serve primarily to demonstrate the functionality of our code, with constraints on the evolution of \HeII\ reionization possible with the advent of future \HeII\ Lyman $\alpha$ programs.

\begin{figure}
\begin{center}
\resizebox{9.5cm}{!}{\includegraphics{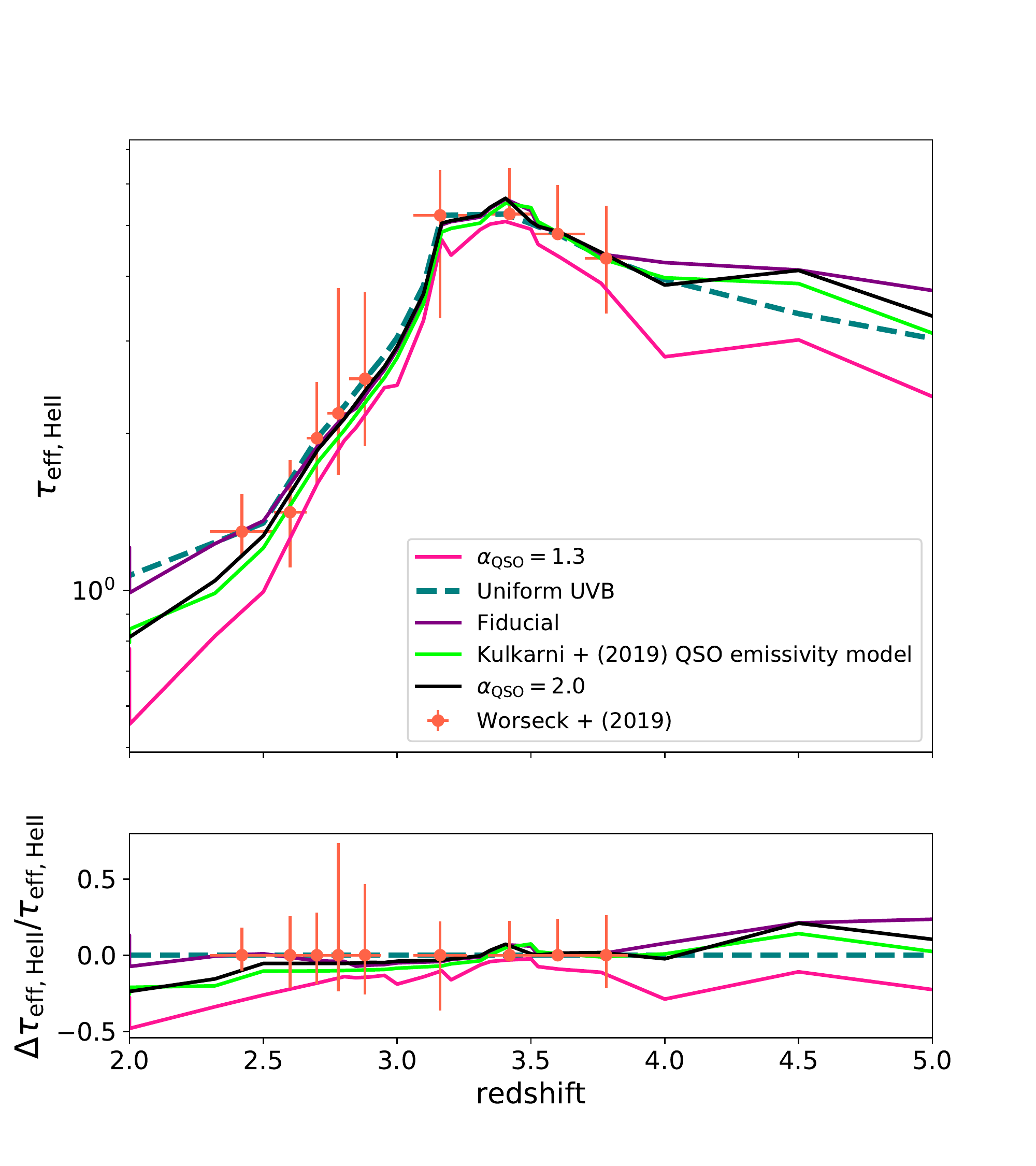}}\\
\end{center}
\caption{The evolution of the effective optical depth of the \HeII\ Lyman $\alpha$ forest for three of our \HeII\ reionization models compared with a model with a uniform UVB (dashed, teal curve). The solid purple curve shows our model with $\alpha_{\rm QSO} = 1.7$, $C_{\rm HeIII } = 3$, $E_{\rm max} = 150$ eV, and $z_{\rm reion,HeII}=2.8-4$. The pink and black curves shows this model, but where $\alpha_{\rm QSO}=1.3$ and $\alpha_{\rm QSO}= 2.0$. The light green curve shows a thermal history based on the quasar emissivity function of \citet{kulkarni19}. The orange error bars show the measured effective optical depths of \citet{worseck19}. The bottom panel shows the relative differences between these four \HeII\ reionization models and the uniform UVB model. 
\label{fig:tauheii}}
\end{figure}

Figure~\ref{fig:tauheii} shows the evolution of the effective optical depth of the \HeII\ Lyman $\alpha$ forest for a selection of our models with our uniform UVB model scaled to the effective optical depth measurements of \citet{worseck19}. The solid curves in the top panel represent our fiducial model (purple), our model with $\alpha_{\rm QSO}=1.3$ (pink), with $\alpha_{\rm QSO}=2.0$ (black), and our model with an ionization fraction based on the quasar emissivity function of \citet{kulkarni19} (green). The dashed teal curve shows our uniform \citet{puchwein19} UVB model. The error bars show the \HeII\ Lyman $\alpha$ forest effective optical depth measurements of \citet{worseck19}. Though our models differ on the order of tens of percent relative to our uniform UVB model, the error bars of the \citet{worseck19} measurements eclipse these differences. However, we estimate that roughly, the size of the errorbars will decrease as the square root of the number of sightlines, and thus with approximately one hundred sightlines (including the 24 sightlines used in the \citealt{worseck19} measurements), future measurements of $\tau_{\rm eff, HeII}$ may be able to distinguish between models with different quasar spectral indices.

\section{Conclusions}
\label{sec:conclusions}

We have constructed a model for \HeII\ reionization in cosmological simulations without the need for in situ radiative transfer. This method includes the thermal impact of the reionization of \HeII\-- capturing the patchiness-- and thus temperature fluctuations and secondary effects such as pressure smoothing of the intergalactic gas. It also includes patchiness in the \HeIII\ field, reflecting the fact that these bubbles develop around biased tracers of the density field.This model is now publicly available as a feature of MP-Gadget, found at: \url{ https://github.com/MP-Gadget/MP-Gadget/blob/master/libgadget/cooling_qso_lightup.c}. The code to produce an input table that specifies \HeII\ reionization parameters can be found at \url{https://github.com/MP-Gadget/MP-Gadget/blob/master/tools/HeII_input_file_maker.py}. 

This tool can be used to investigate the parameter space of \HeII\ reionization and frees future simulations from the rigid \HeII\ parameters in UVB models. Because of its flexibility and efficiency, our model can perform a Bayesian analysis of \HeII\ reionization, something previously unattainable with \HeII\ reionization simulations that required radiative transfer.

Our simulations show thermal histories consistent with historical temperature measurements of the IGM-- though uniform UVB models also are able to reproduce the thermal evolution at mean cosmic density. However, the thermal evolution of the gas in our simulations more accurately reflects the physics of \HeII\ reionization than simulations that use a uniform UVB to model the \HeII\ reionization photoheating and captures the spatial variance in temperature from a patchy reionization. We show the evolution in the dispersion of the temperature-density relation with patchy reionization, finding that the scatter at redshifts near the conclusion of \HeII\ reioniziation is consistent with previous radiative transfer simulations \citep{mcquinn09,compostella13,compostella14,laplante17,laplante18}. 

Though the one-dimensional statistics of the Lyman $\alpha$ forest, specifically the one-dimensional flux power spectrum, reflect a modest change between models with and without patchy \HeII\ reionization as previously shown in \citet{mcquinn11} and \citet{laplante17}, we find signatures of inhomogeneous \HeII\ reionization at the level of several percent that are mildly favored by the measurements of \citet{walther18} and \citet{irsic17}. With precision cosmology requiring percent-level accuracy for models of the flux power spectrum, it may be necessary to marginalize over the parameters of \HeII\ reionization to improve constraints. 

The recent debut of new \HeII\ Lyman $\alpha$ forest sightlines require large \HeII\ reionization simulations for future progress in observationally constraining \HeII\ reionization. The \HeII\ Lyman $\alpha$ forest extracted from our simulations can eventually serve as a theoretical tool to better constrain the evolution and conclusion of \HeII\ reionization. While the small number of sightlines limit this progress, flexible \HeII\ reionization simulations will be crucial for future \HeII\ Lyman $\alpha$ programs.

Until now, including \HeII\ reionization in hydrodynamic simulations required computationally expensive radiative transfer. Going forward, hydrodynamic simulations can include \HeII\ reionization with a minimal increase in simulation time. The addition of the long-neglected physics of \HeII\ reionization and the ability to customize the \HeII\ reionization parameters should incentivize future work on \HeII\ reionization from both the observational and theoretical perspective.

\section*{Acknowledgments}

We thank the anonymous referee for their thoughtful comments and suggestions. We thank Matt McQuinn, Fred Davies, and Anson D'Aloisio for useful discussions. This work was supported by NSF award AST-1817256. Computing resources were provided by NSF XSEDE allocation AST180058. The authors acknowledge the Texas Advanced Computing Center (TACC) at The University of Texas at Austin for providing HPC resources on the Frontera supercomputer that have contributed to the research results reported within this paper, via grant NSF AST-1616168 and allocation FTA-DiMatteo.

\section*{Data Availability}
The simulation data underlying this article can be reproduced using the public code at \url{ https://github.com/MP-Gadget/MP-Gadget}.

\bibliographystyle{mnras}
\bibliography{References}

\begin{thebibliography}{}
\makeatletter
\relax
\def\mn@urlcharsother{\let\do\@makeother \do\$\do\&\do\#\do\^\do\_\do\%\do\~}
\def\mn@doi{\begingroup\mn@urlcharsother \@ifnextchar [ {\mn@doi@}
  {\mn@doi@[]}}
\def\mn@doi@[#1]#2{\def\@tempa{#1}\ifx\@tempa\@empty \href
  {http://dx.doi.org/#2} {doi:#2}\else \href {http://dx.doi.org/#2} {#1}\fi
  \endgroup}
\def\mn@eprint#1#2{\mn@eprint@#1:#2::\@nil}
\def\mn@eprint@arXiv#1{\href {http://arxiv.org/abs/#1} {{\tt arXiv:#1}}}
\def\mn@eprint@dblp#1{\href {http://dblp.uni-trier.de/rec/bibtex/#1.xml}
  {dblp:#1}}
\def\mn@eprint@#1:#2:#3:#4\@nil{\def\@tempa {#1}\def\@tempb {#2}\def\@tempc
  {#3}\ifx \@tempc \@empty \let \@tempc \@tempb \let \@tempb \@tempa \fi \ifx
  \@tempb \@empty \def\@tempb {arXiv}\fi \@ifundefined
  {mn@eprint@\@tempb}{\@tempb:\@tempc}{\expandafter \expandafter \csname
  mn@eprint@\@tempb\endcsname \expandafter{\@tempc}}}

\bibitem[\protect\citeauthoryear{{Becker}, {Bolton}, {Haehnelt}  \&
  {Sargent}}{{Becker} et~al.}{2011}]{becker11}
{Becker} G.~D.,  {Bolton} J.~S.,  {Haehnelt} M.~G.,   {Sargent} W.~L.~W.,
  2011, \mn@doi [\mnras] {10.1111/j.1365-2966.2010.17507.x}, \href
  {http://adsabs.harvard.edu/abs/2011MNRAS.410.1096B} {410, 1096}

\bibitem[\protect\citeauthoryear{{Boera}, {Murphy}, {Becker}  \&
  {Bolton}}{{Boera} et~al.}{2014}]{boera14}
{Boera} E.,  {Murphy} M.~T.,  {Becker} G.~D.,   {Bolton} J.~S.,  2014, \mn@doi
  [\mnras] {10.1093/mnras/stu660}, \href
  {http://adsabs.harvard.edu/abs/2014MNRAS.441.1916B} {441, 1916}

\bibitem[\protect\citeauthoryear{{Boera}, {Becker}, {Bolton}  \&
  {Nasir}}{{Boera} et~al.}{2019}]{boera19}
{Boera} E.,  {Becker} G.~D.,  {Bolton} J.~S.,   {Nasir} F.,  2019, \mn@doi
  [\apj] {10.3847/1538-4357/aafee4}, \href
  {https://ui.adsabs.harvard.edu/abs/2019ApJ...872..101B} {872, 101}

\bibitem[\protect\citeauthoryear{{Bolton}, {Becker}, {Haehnelt}  \&
  {Viel}}{{Bolton} et~al.}{2014}]{bolton14}
{Bolton} J.~S.,  {Becker} G.~D.,  {Haehnelt} M.~G.,   {Viel} M.,  2014, \mn@doi
  [\mnras] {10.1093/mnras/stt2374}, \href
  {http://adsabs.harvard.edu/abs/2014MNRAS.438.2499B} {438, 2499}

\bibitem[\protect\citeauthoryear{{Calura}, {Tescari}, {D'Odorico}, {Viel},
  {Cristiani}, {Kim}  \& {Bolton}}{{Calura} et~al.}{2012}]{calura12}
{Calura} F.,  {Tescari} E.,  {D'Odorico} V.,  {Viel} M.,  {Cristiani} S.,
  {Kim} T.~S.,   {Bolton} J.~S.,  2012, \mn@doi [\mnras]
  {10.1111/j.1365-2966.2012.20811.x}, \href
  {https://ui.adsabs.harvard.edu/abs/2012MNRAS.422.3019C} {422, 3019}

\bibitem[\protect\citeauthoryear{{Compostella}, {Cantalupo}  \&
  {Porciani}}{{Compostella} et~al.}{2013}]{compostella13}
{Compostella} M.,  {Cantalupo} S.,   {Porciani} C.,  2013, \mn@doi [\mnras]
  {10.1093/mnras/stt1510}, \href
  {https://ui.adsabs.harvard.edu/abs/2013MNRAS.435.3169C} {435, 3169}

\bibitem[\protect\citeauthoryear{{Compostella}, {Cantalupo}  \&
  {Porciani}}{{Compostella} et~al.}{2014}]{compostella14}
{Compostella} M.,  {Cantalupo} S.,   {Porciani} C.,  2014, \mn@doi [\mnras]
  {10.1093/mnras/stu2035}, \href
  {https://ui.adsabs.harvard.edu/abs/2014MNRAS.445.4186C} {445, 4186}

\bibitem[\protect\citeauthoryear{{Fang} \& {White}}{{Fang} \&
  {White}}{2004}]{fang04}
{Fang} T.,  {White} M.,  2004, \mn@doi [\apjl] {10.1086/420965}, \href
  {https://ui.adsabs.harvard.edu/abs/2004ApJ...606L...9F} {606, L9}

\bibitem[\protect\citeauthoryear{{Faucher-Giguere} \& {-A.}}{{Faucher-Giguere}
  \& {-A.}}{2019}]{fg19}
{Faucher-Giguere} {-A.} C.,  2019, arXiv e-prints, \href
  {https://ui.adsabs.harvard.edu/abs/2019arXiv190308657F} {p. arXiv:1903.08657}

\bibitem[\protect\citeauthoryear{{Feng}, {Di-Matteo}, {Croft}, {Bird},
  {Battaglia}  \& {Wilkins}}{{Feng} et~al.}{2016}]{feng16}
{Feng} Y.,  {Di-Matteo} T.,  {Croft} R.~A.,  {Bird} S.,  {Battaglia} N.,
  {Wilkins} S.,  2016, \mn@doi [\mnras] {10.1093/mnras/stv2484}, \href
  {https://ui.adsabs.harvard.edu/abs/2016MNRAS.455.2778F} {455, 2778}

\bibitem[\protect\citeauthoryear{Feng, Bird, Anderson, Font-Ribera  \&
  Pedersen}{Feng et~al.}{2018}]{yu_feng_2018_1451799}
Feng Y.,  Bird S.,  Anderson L.,  Font-Ribera A.,   Pedersen C.,  2018,
  MP-Gadget/MP-Gadget: A tag for getting a DOI,
  \mn@doi{10.5281/zenodo.1451799}, \url
  {https://doi.org/10.5281/zenodo.1451799}

\bibitem[\protect\citeauthoryear{{Furlanetto} \& {Oh}}{{Furlanetto} \&
  {Oh}}{2008}]{furlanetto08}
{Furlanetto} S.~R.,  {Oh} S.~P.,  2008, \mn@doi [\apj] {10.1086/588546}, \href
  {https://ui.adsabs.harvard.edu/abs/2008ApJ...681....1F} {681, 1}

\bibitem[\protect\citeauthoryear{{Gaikwad} et~al.,}{{Gaikwad}
  et~al.}{2020}]{gaikwad20}
{Gaikwad} P.,  et~al., 2020, arXiv e-prints, \href
  {https://ui.adsabs.harvard.edu/abs/2020arXiv200110018G} {p. arXiv:2001.10018}

\bibitem[\protect\citeauthoryear{{Garzilli}, {Bolton}, {Kim}, {Leach}  \&
  {Viel}}{{Garzilli} et~al.}{2012}]{garzilli12}
{Garzilli} A.,  {Bolton} J.~S.,  {Kim} T.-S.,  {Leach} S.,   {Viel} M.,  2012,
  \mn@doi [\mnras] {10.1111/j.1365-2966.2012.21223.x}, \href
  {http://adsabs.harvard.edu/abs/2012MNRAS.424.1723G} {424, 1723}

\bibitem[\protect\citeauthoryear{{Gnedin} \& {Hui}}{{Gnedin} \&
  {Hui}}{1998}]{gnedin98}
{Gnedin} N.~Y.,  {Hui} L.,  1998, \mn@doi [\mnras]
  {10.1046/j.1365-8711.1998.01249.x}, \href
  {https://ui.adsabs.harvard.edu/abs/1998MNRAS.296...44G} {296, 44}

\bibitem[\protect\citeauthoryear{{Haardt} \& {Madau}}{{Haardt} \&
  {Madau}}{2012}]{haardt12}
{Haardt} F.,  {Madau} P.,  2012, \mn@doi [\apj] {10.1088/0004-637X/746/2/125},
  \href {https://ui.adsabs.harvard.edu/abs/2012ApJ...746..125H} {746, 125}

\bibitem[\protect\citeauthoryear{{Heap}, {Williger}, {Smette}, {Hubeny},
  {Sahu}, {Jenkins}, {Tripp}  \& {Winkler}}{{Heap} et~al.}{2000}]{heap00}
{Heap} S.~R.,  {Williger} G.~M.,  {Smette} A.,  {Hubeny} I.,  {Sahu} M.~S.,
  {Jenkins} E.~B.,  {Tripp} T.~M.,   {Winkler} J.~N.,  2000, \mn@doi [\apj]
  {10.1086/308719}, \href
  {https://ui.adsabs.harvard.edu/abs/2000ApJ...534...69H} {534, 69}

\bibitem[\protect\citeauthoryear{{Hiss}, {Walther}, {Hennawi}, {O{\~n}orbe},
  {O'Meara}, {Rorai}  \& {Luki{\'c}}}{{Hiss} et~al.}{2018}]{hiss18}
{Hiss} H.,  {Walther} M.,  {Hennawi} J.~F.,  {O{\~n}orbe} J.,  {O'Meara} J.~M.,
   {Rorai} A.,   {Luki{\'c}} Z.,  2018, \mn@doi [\apj]
  {10.3847/1538-4357/aada86}, \href
  {https://ui.adsabs.harvard.edu/abs/2018ApJ...865...42H} {865, 42}

\bibitem[\protect\citeauthoryear{{Hogan}, {Anderson}  \& {Rugers}}{{Hogan}
  et~al.}{1997}]{hogan97}
{Hogan} C.~J.,  {Anderson} S.~F.,   {Rugers} M.~H.,  1997, \mn@doi [\aj]
  {10.1086/118366}, \href
  {https://ui.adsabs.harvard.edu/abs/1997AJ....113.1495H} {113, 1495}

\bibitem[\protect\citeauthoryear{{Hui} \& {Gnedin}}{{Hui} \&
  {Gnedin}}{1997}]{gnedin97}
{Hui} L.,  {Gnedin} N.~Y.,  1997, \mn@doi [\mnras] {10.1093/mnras/292.1.27},
  \href {https://ui.adsabs.harvard.edu/abs/1997MNRAS.292...27H} {292, 27}

\bibitem[\protect\citeauthoryear{{Hui} \& {Haiman}}{{Hui} \&
  {Haiman}}{2003}]{hui03}
{Hui} L.,  {Haiman} Z.,  2003, \mn@doi [\apj] {10.1086/377229}, \href
  {https://ui.adsabs.harvard.edu/abs/2003ApJ...596....9H} {596, 9}

\bibitem[\protect\citeauthoryear{{Ir{\v{s}}i{\v{c}}}
  et~al.,}{{Ir{\v{s}}i{\v{c}}} et~al.}{2017}]{irsic17}
{Ir{\v{s}}i{\v{c}}} V.,  et~al., 2017, \mn@doi [\mnras]
  {10.1093/mnras/stw3372}, \href
  {https://ui.adsabs.harvard.edu/abs/2017MNRAS.466.4332I} {466, 4332}

\bibitem[\protect\citeauthoryear{{Khaire}}{{Khaire}}{2017}]{khaire17}
{Khaire} V.,  2017, \mn@doi [\mnras] {10.1093/mnras/stx1487}, \href
  {https://ui.adsabs.harvard.edu/abs/2017MNRAS.471..255K} {471, 255}

\bibitem[\protect\citeauthoryear{{Khaire} \& {Srianand}}{{Khaire} \&
  {Srianand}}{2015}]{khaire15}
{Khaire} V.,  {Srianand} R.,  2015, \mn@doi [\mnras] {10.1093/mnrasl/slv060},
  \href {https://ui.adsabs.harvard.edu/abs/2015MNRAS.451L..30K} {451, L30}

\bibitem[\protect\citeauthoryear{{Kim}, {Bolton}, {Viel}, {Haehnelt}  \&
  {Carswell}}{{Kim} et~al.}{2007}]{kim07}
{Kim} T.~S.,  {Bolton} J.~S.,  {Viel} M.,  {Haehnelt} M.~G.,   {Carswell}
  R.~F.,  2007, \mn@doi [\mnras] {10.1111/j.1365-2966.2007.12406.x}, \href
  {https://ui.adsabs.harvard.edu/abs/2007MNRAS.382.1657K} {382, 1657}

\bibitem[\protect\citeauthoryear{{Kulkarni}, {Worseck}  \&
  {Hennawi}}{{Kulkarni} et~al.}{2019}]{kulkarni19}
{Kulkarni} G.,  {Worseck} G.,   {Hennawi} J.~F.,  2019, \mn@doi [\mnras]
  {10.1093/mnras/stz1493}, \href
  {https://ui.adsabs.harvard.edu/abs/2019MNRAS.488.1035K} {488, 1035}

\bibitem[\protect\citeauthoryear{{La Plante}, {Trac}, {Croft}  \& {Cen}}{{La
  Plante} et~al.}{2017}]{laplante17}
{La Plante} P.,  {Trac} H.,  {Croft} R.,   {Cen} R.,  2017, \mn@doi [\apj]
  {10.3847/1538-4357/aa7136}, \href
  {https://ui.adsabs.harvard.edu/abs/2017ApJ...841...87L} {841, 87}

\bibitem[\protect\citeauthoryear{{La Plante}, {Trac}, {Croft}  \& {Cen}}{{La
  Plante} et~al.}{2018}]{laplante18}
{La Plante} P.,  {Trac} H.,  {Croft} R.,   {Cen} R.,  2018, \mn@doi [\apj]
  {10.3847/1538-4357/aae693}, \href
  {https://ui.adsabs.harvard.edu/abs/2018ApJ...868..106L} {868, 106}

\bibitem[\protect\citeauthoryear{{Lai}, {Lidz}, {Hernquist}  \&
  {Zaldarriaga}}{{Lai} et~al.}{2006}]{lai06}
{Lai} K.,  {Lidz} A.,  {Hernquist} L.,   {Zaldarriaga} M.,  2006, \mn@doi
  [\apj] {10.1086/503320}, \href
  {https://ui.adsabs.harvard.edu/abs/2006ApJ...644...61L} {644, 61}

\bibitem[\protect\citeauthoryear{{Lidz}, {Faucher-Gigu{\`e}re}, {Dall'Aglio},
  {McQuinn}, {Fechner}, {Zaldarriaga}, {Hernquist}  \& {Dutta}}{{Lidz}
  et~al.}{2010}]{lidz10}
{Lidz} A.,  {Faucher-Gigu{\`e}re} C.-A.,  {Dall'Aglio} A.,  {McQuinn} M.,
  {Fechner} C.,  {Zaldarriaga} M.,  {Hernquist} L.,   {Dutta} S.,  2010,
  \mn@doi [\apj] {10.1088/0004-637X/718/1/199}, \href
  {http://adsabs.harvard.edu/abs/2010ApJ...718..199L} {718, 199}

\bibitem[\protect\citeauthoryear{{Lusso}, {Worseck}, {Hennawi}, {Prochaska},
  {Vignali}, {Stern}  \& {O'Meara}}{{Lusso} et~al.}{2015}]{lusso15}
{Lusso} E.,  {Worseck} G.,  {Hennawi} J.~F.,  {Prochaska} J.~X.,  {Vignali} C.,
   {Stern} J.,   {O'Meara} J.~M.,  2015, \mn@doi [\mnras]
  {10.1093/mnras/stv516}, \href
  {http://adsabs.harvard.edu/abs/2015MNRAS.449.4204L} {449, 4204}

\bibitem[\protect\citeauthoryear{{Madau} \& {Meiksin}}{{Madau} \&
  {Meiksin}}{1994}]{madau94}
{Madau} P.,  {Meiksin} A.,  1994, \mn@doi [\apjl] {10.1086/187546}, \href
  {https://ui.adsabs.harvard.edu/abs/1994ApJ...433L..53M} {433, L53}

\bibitem[\protect\citeauthoryear{{Madau}, {Haardt}  \& {Rees}}{{Madau}
  et~al.}{1999}]{madau99}
{Madau} P.,  {Haardt} F.,   {Rees} M.~J.,  1999, \mn@doi [\apj]
  {10.1086/306975}, \href
  {https://ui.adsabs.harvard.edu/abs/1999ApJ...514..648M} {514, 648}

\bibitem[\protect\citeauthoryear{{McQuinn}}{{McQuinn}}{2009}]{mcquinnGP}
{McQuinn} M.,  2009, \mn@doi [\apjl] {10.1088/0004-637X/704/2/L89}, \href
  {http://adsabs.harvard.edu/abs/2009ApJ...704L..89M} {704, L89}

\bibitem[\protect\citeauthoryear{{McQuinn} \& {Upton Sanderbeck}}{{McQuinn} \&
  {Upton Sanderbeck}}{2016}]{mcquinn16}
{McQuinn} M.,  {Upton Sanderbeck} P.~R.,  2016, \mn@doi [\mnras]
  {10.1093/mnras/stv2675}, \href
  {https://ui.adsabs.harvard.edu/abs/2016MNRAS.456...47M} {456, 47}

\bibitem[\protect\citeauthoryear{{McQuinn}, {Lidz}, {Zaldarriaga}, {Hernquist},
  {Hopkins}, {Dutta}  \& {Faucher-Gigu{\`e}re}}{{McQuinn}
  et~al.}{2009}]{mcquinn09}
{McQuinn} M.,  {Lidz} A.,  {Zaldarriaga} M.,  {Hernquist} L.,  {Hopkins} P.~F.,
   {Dutta} S.,   {Faucher-Gigu{\`e}re} C.-A.,  2009, \mn@doi [\apj]
  {10.1088/0004-637X/694/2/842}, \href
  {https://ui.adsabs.harvard.edu/abs/2009ApJ...694..842M} {694, 842}

\bibitem[\protect\citeauthoryear{{McQuinn}, {Hernquist}, {Lidz}  \&
  {Zaldarriaga}}{{McQuinn} et~al.}{2011}]{mcquinn11}
{McQuinn} M.,  {Hernquist} L.,  {Lidz} A.,   {Zaldarriaga} M.,  2011, \mn@doi
  [\mnras] {10.1111/j.1365-2966.2011.18788.x}, \href
  {https://ui.adsabs.harvard.edu/abs/2011MNRAS.415..977M} {415, 977}

\bibitem[\protect\citeauthoryear{{Meiksin}}{{Meiksin}}{2000}]{meiksin00}
{Meiksin} A.,  2000, \mn@doi [\mnras] {10.1046/j.1365-8711.2000.03315.x}, \href
  {https://ui.adsabs.harvard.edu/abs/2000MNRAS.314..566M} {314, 566}

\bibitem[\protect\citeauthoryear{{Miralda-Escude}}{{Miralda-Escude}}{1993}]{miraldaescude93}
{Miralda-Escude} J.,  1993, \mn@doi [\mnras] {10.1093/mnras/262.1.273}, \href
  {https://ui.adsabs.harvard.edu/abs/1993MNRAS.262..273M} {262, 273}

\bibitem[\protect\citeauthoryear{{Miralda-Escud{\'e}} \&
  {Rees}}{{Miralda-Escud{\'e}} \& {Rees}}{1994}]{miraldaescude94}
{Miralda-Escud{\'e}} J.,  {Rees} M.~J.,  1994, \mn@doi [\mnras]
  {10.1093/mnras/266.2.343}, \href
  {https://ui.adsabs.harvard.edu/abs/1994MNRAS.266..343M} {266, 343}

\bibitem[\protect\citeauthoryear{{Miralda-Escud{\'e}}, {Haehnelt}  \&
  {Rees}}{{Miralda-Escud{\'e}} et~al.}{2000}]{miraldaescude00}
{Miralda-Escud{\'e}} J.,  {Haehnelt} M.,   {Rees} M.~J.,  2000, \mn@doi [\apj]
  {10.1086/308330}, \href
  {https://ui.adsabs.harvard.edu/abs/2000ApJ...530....1M} {530, 1}

\bibitem[\protect\citeauthoryear{{O{\~n}orbe}, {Davies}, {Luki{\'c}}, {},
  {Hennawi}  \& {Sorini}}{{O{\~n}orbe} et~al.}{2019}]{onorbe19}
{O{\~n}orbe} J.,  {Davies} F.~B.,  {Luki{\'c}} {} Z.,  {Hennawi} J.~F.,
  {Sorini} D.,  2019, \mn@doi [\mnras] {10.1093/mnras/stz984}, \href
  {https://ui.adsabs.harvard.edu/abs/2019MNRAS.486.4075O} {486, 4075}

\bibitem[\protect\citeauthoryear{{Puchwein}, {Bolton}, {Haehnelt}, {Madau},
  {Becker}  \& {Haardt}}{{Puchwein} et~al.}{2015}]{puchwein15}
{Puchwein} E.,  {Bolton} J.~S.,  {Haehnelt} M.~G.,  {Madau} P.,  {Becker}
  G.~D.,   {Haardt} F.,  2015, \mn@doi [\mnras] {10.1093/mnras/stv773}, \href
  {https://ui.adsabs.harvard.edu/abs/2015MNRAS.450.4081P} {450, 4081}

\bibitem[\protect\citeauthoryear{{Puchwein}, {Haardt}, {Haehnelt}  \&
  {Madau}}{{Puchwein} et~al.}{2019}]{puchwein19}
{Puchwein} E.,  {Haardt} F.,  {Haehnelt} M.~G.,   {Madau} P.,  2019, \mn@doi
  [\mnras] {10.1093/mnras/stz222}, \href
  {https://ui.adsabs.harvard.edu/abs/2019MNRAS.485...47P} {485, 47}

\bibitem[\protect\citeauthoryear{{Reimers}, {Kohler}, {Wisotzki}, {Groote},
  {Rodriguez-Pascual}  \& {Wamsteker}}{{Reimers} et~al.}{1997}]{reimers97}
{Reimers} D.,  {Kohler} S.,  {Wisotzki} L.,  {Groote} D.,  {Rodriguez-Pascual}
  P.,   {Wamsteker} W.,  1997, \aap, \href
  {https://ui.adsabs.harvard.edu/abs/1997A&A...327..890R} {327, 890}

\bibitem[\protect\citeauthoryear{{Schaye}, {Theuns}, {Rauch}, {Efstathiou}  \&
  {Sargent}}{{Schaye} et~al.}{2000}]{schaye00}
{Schaye} J.,  {Theuns} T.,  {Rauch} M.,  {Efstathiou} G.,   {Sargent} W.~L.~W.,
   2000, \mn@doi [\mnras] {10.1046/j.1365-8711.2000.03815.x}, \href
  {http://adsabs.harvard.edu/abs/2000MNRAS.318..817S} {318, 817}

\bibitem[\protect\citeauthoryear{{Schaye}, {Aguirre}, {Kim}, {Theuns}, {Rauch}
  \& {Sargent}}{{Schaye} et~al.}{2003}]{schaye03}
{Schaye} J.,  {Aguirre} A.,  {Kim} T.-S.,  {Theuns} T.,  {Rauch} M.,
  {Sargent} W. L.~W.,  2003, \mn@doi [\apj] {10.1086/378044}, \href
  {https://ui.adsabs.harvard.edu/abs/2003ApJ...596..768S} {596, 768}

\bibitem[\protect\citeauthoryear{{Shull}, {France}, {Danforth}, {Smith}  \&
  {Tumlinson}}{{Shull} et~al.}{2010}]{shull10}
{Shull} M.,  {France} K.,  {Danforth} C.,  {Smith} B.,   {Tumlinson} J.,  2010,
  arxiv:1008.2957, \href {http://adsabs.harvard.edu/abs/2010arXiv1008.2957S} {}

\bibitem[\protect\citeauthoryear{{Shull}, {Stevans}  \& {Danforth}}{{Shull}
  et~al.}{2012}]{shull12}
{Shull} J.~M.,  {Stevans} M.,   {Danforth} C.~W.,  2012, \mn@doi [\apj]
  {10.1088/0004-637X/752/2/162}, \href
  {https://ui.adsabs.harvard.edu/abs/2012ApJ...752..162S} {752, 162}

\bibitem[\protect\citeauthoryear{{Springel}}{{Springel}}{2005}]{Springel:2005}
{Springel} V.,  2005, \mnras, \href
  {http://adsabs.harvard.edu/abs/2003MNRAS.339..289S} {364, 1105}

\bibitem[\protect\citeauthoryear{{Stevans}, {Shull}, {Danforth}  \&
  {Tilton}}{{Stevans} et~al.}{2014}]{stevans14}
{Stevans} M.~L.,  {Shull} J.~M.,  {Danforth} C.~W.,   {Tilton} E.~M.,  2014,
  \mn@doi [\apj] {10.1088/0004-637X/794/1/75}, \href
  {https://ui.adsabs.harvard.edu/abs/2014ApJ...794...75S} {794, 75}

\bibitem[\protect\citeauthoryear{{Syphers} \& {Shull}}{{Syphers} \&
  {Shull}}{2013}]{syphers13}
{Syphers} D.,  {Shull} J.~M.,  2013, \mn@doi [\apj]
  {10.1088/0004-637X/765/2/119}, \href
  {https://ui.adsabs.harvard.edu/abs/2013ApJ...765..119S} {765, 119}

\bibitem[\protect\citeauthoryear{{Syphers} \& {Shull}}{{Syphers} \&
  {Shull}}{2014}]{syphers14}
{Syphers} D.,  {Shull} J.~M.,  2014, \mn@doi [\apj]
  {10.1088/0004-637X/784/1/42}, \href
  {https://ui.adsabs.harvard.edu/abs/2014ApJ...784...42S} {784, 42}

\bibitem[\protect\citeauthoryear{{Telfer}, {Zheng}, {Kriss}  \&
  {Davidsen}}{{Telfer} et~al.}{2002}]{telfer02}
{Telfer} R.~C.,  {Zheng} W.,  {Kriss} G.~A.,   {Davidsen} A.~F.,  2002, \apj,
  565, 773

\bibitem[\protect\citeauthoryear{{Theuns} \& {Zaroubi}}{{Theuns} \&
  {Zaroubi}}{2000}]{theuns00}
{Theuns} T.,  {Zaroubi} S.,  2000, \mn@doi [\mnras]
  {10.1046/j.1365-8711.2000.03729.x}, \href
  {https://ui.adsabs.harvard.edu/abs/2000MNRAS.317..989T} {317, 989}

\bibitem[\protect\citeauthoryear{{Tittley} \& {Meiksin}}{{Tittley} \&
  {Meiksin}}{2007}]{tittley07}
{Tittley} E.~R.,  {Meiksin} A.,  2007, \mn@doi [\mnras]
  {10.1111/j.1365-2966.2007.12214.x}, \href
  {https://ui.adsabs.harvard.edu/abs/2007MNRAS.380.1369T} {380, 1369}

\bibitem[\protect\citeauthoryear{{Upton Sanderbeck}, {D'Aloisio}  \&
  {McQuinn}}{{Upton Sanderbeck} et~al.}{2016}]{uptonsanderbeck16}
{Upton Sanderbeck} P.~R.,  {D'Aloisio} A.,   {McQuinn} M.~J.,  2016, \mn@doi
  [\mnras] {10.1093/mnras/stw1117}, \href
  {https://ui.adsabs.harvard.edu/abs/2016MNRAS.460.1885U} {460, 1885}

\bibitem[\protect\citeauthoryear{{Upton Sanderbeck}, {McQuinn}, {D'Aloisio}  \&
  {Werk}}{{Upton Sanderbeck} et~al.}{2018}]{uptonsanderbeck18}
{Upton Sanderbeck} P.~R.,  {McQuinn} M.,  {D'Aloisio} A.,   {Werk} J.~K.,
  2018, \mn@doi [\apj] {10.3847/1538-4357/aaeff2}, \href
  {https://ui.adsabs.harvard.edu/abs/2018ApJ...869..159U} {869, 159}

\bibitem[\protect\citeauthoryear{{Walther}, {Hennawi}, {Hiss}, {O{\~n}orbe},
  {Lee}, {Rorai}  \& {O'Meara}}{{Walther} et~al.}{2018}]{walther18}
{Walther} M.,  {Hennawi} J.~F.,  {Hiss} H.,  {O{\~n}orbe} J.,  {Lee} K.-G.,
  {Rorai} A.,   {O'Meara} J.,  2018, \mn@doi [\apj] {10.3847/1538-4357/aa9c81},
  \href {https://ui.adsabs.harvard.edu/abs/2018ApJ...852...22W} {852, 22}

\bibitem[\protect\citeauthoryear{{Walther}, {O{\~n}orbe}, {Hennawi}  \&
  {Luki{\'c}}}{{Walther} et~al.}{2019}]{walther19}
{Walther} M.,  {O{\~n}orbe} J.,  {Hennawi} J.~F.,   {Luki{\'c}} Z.,  2019,
  \mn@doi [\apj] {10.3847/1538-4357/aafad1}, \href
  {https://ui.adsabs.harvard.edu/abs/2019ApJ...872...13W} {872, 13}

\bibitem[\protect\citeauthoryear{{White}, {Pope}, {Carlson}, {Heitmann},
  {Habib}, {Fasel}, {Daniel}  \& {Lukic}}{{White} et~al.}{2010}]{white10}
{White} M.,  {Pope} A.,  {Carlson} J.,  {Heitmann} K.,  {Habib} S.,  {Fasel}
  P.,  {Daniel} D.,   {Lukic} Z.,  2010, \mn@doi [\apj]
  {10.1088/0004-637X/713/1/383}, \href
  {https://ui.adsabs.harvard.edu/abs/2010ApJ...713..383W} {713, 383}

\bibitem[\protect\citeauthoryear{{Worseck} et~al.,}{{Worseck}
  et~al.}{2011}]{worseck11}
{Worseck} G.,  et~al., 2011, \mn@doi [\apjl] {10.1088/2041-8205/733/2/L24},
  \href {http://adsabs.harvard.edu/abs/2011ApJ...733L..24W} {733, L24}

\bibitem[\protect\citeauthoryear{{Worseck}, {Prochaska}, {Hennawi}  \&
  {McQuinn}}{{Worseck} et~al.}{2016}]{worseck16}
{Worseck} G.,  {Prochaska} J.~X.,  {Hennawi} J.~F.,   {McQuinn} M.,  2016,
  \mn@doi [\apj] {10.3847/0004-637X/825/2/144}, \href
  {http://adsabs.harvard.edu/abs/2016ApJ...825..144W} {825, 144}

\bibitem[\protect\citeauthoryear{{Worseck}, {Davies}, {Hennawi}  \&
  {Prochaska}}{{Worseck} et~al.}{2019}]{worseck19}
{Worseck} G.,  {Davies} F.~B.,  {Hennawi} J.~F.,   {Prochaska} J.~X.,  2019,
  \mn@doi [\apj] {10.3847/1538-4357/ab0fa1}, \href
  {https://ui.adsabs.harvard.edu/abs/2019ApJ...875..111W} {875, 111}

\bibitem[\protect\citeauthoryear{{Wyithe} \& {Loeb}}{{Wyithe} \&
  {Loeb}}{2003}]{wyithe03}
{Wyithe} J. S.~B.,  {Loeb} A.,  2003, \mn@doi [\apj] {10.1086/367721}, \href
  {https://ui.adsabs.harvard.edu/abs/2003ApJ...586..693W} {586, 693}

\bibitem[\protect\citeauthoryear{{Zaldarriaga}}{{Zaldarriaga}}{2002}]{zaldarriaga02}
{Zaldarriaga} M.,  2002, \mn@doi [\apj] {10.1086/324212}, \href
  {https://ui.adsabs.harvard.edu/abs/2002ApJ...564..153Z} {564, 153}

\bibitem[\protect\citeauthoryear{{Zheng} et~al.,}{{Zheng}
  et~al.}{2004}]{zheng04}
{Zheng} W.,  et~al., 2004, \mn@doi [\apj] {10.1086/382498}, \href
  {https://ui.adsabs.harvard.edu/abs/2004ApJ...605..631Z} {605, 631}

\bibitem[\protect\citeauthoryear{{Zheng}, {Syphers}, {Meiksin}, {Kriss},
  {Schneider}, {York}  \& {Anderson}}{{Zheng} et~al.}{2015}]{zheng15}
{Zheng} W.,  {Syphers} D.,  {Meiksin} A.,  {Kriss} G.~A.,  {Schneider} D.~P.,
  {York} D.~G.,   {Anderson} S.~F.,  2015, \mn@doi [\apj]
  {10.1088/0004-637X/806/1/142}, \href
  {https://ui.adsabs.harvard.edu/abs/2015ApJ...806..142Z} {806, 142}

\makeatother
\end{thebibliography}

\bsp	
\label{lastpage}

\end{document}